\documentclass{aa}  
\usepackage{graphicx}
\usepackage{txfonts}
\newcommand{\mincir}{\raise -2.truept\hbox{\rlap{\hbox{$\sim$}}\raise5.truept
\hbox{$<$}\ }}
\newcommand{\magcir}{\raise -2.truept\hbox{\rlap{\hbox{$\sim$}}\raise5.truept
\hbox{$>$}\ }}
\newcommand{\siml}{\raise -2.truept\hbox{\rlap{\hbox{$\sim$}}\raise5.truept
\hbox{$<$}\ }}
\newcommand{\simg}{\raise -2.truept\hbox{\rlap{\hbox{$\sim$}}\raise5.truept
\hbox{$>$}\ }}
\newcommand{\be}{\begin{equation}}
\newcommand{\ee}{\end{equation}}
\newcommand{\ba}{\begin{eqnarray}}
\newcommand{\ea}{\end{eqnarray}}
\newcommand {\kpc} {$\mathrm{h_{70}^{-1}}$ kpc $\;$}

\newcommand {\h} {$h_{70}^{-1}$ Mpc$\;$}
\newcommand {\hh} {$h_{70}^{-1}$ Mpc}
\newcommand {\hhh} {\;h_{70}^{-1} \mathrm{Mpc}}
\newcommand {\ks} {km~s$^{-1} \;$}
\newcommand {\kss} {km~s$^{-1}$}

\newcommand {\mqua} {$\times 10^{14}\;h_{70}^{-1}\;M_{\odot} \;$}
\newcommand {\mquaa} {$\times 10^{14}\;h_{70}^{-1}\;M_{\odot}$}
\newcommand {\mqui} {$\times 10^{15}\;h_{70}^{-1}\;M_{\odot} \;$}
\newcommand {\mquii} {$\times 10^{15}\;h_{70}^{-1}\;M_{\odot}$}

\newcommand{\degree}{\ensuremath{\mathrm{^\circ}}}
\newcommand{\arcm}{\ensuremath{\mathrm{^\prime}\;}}

\newcommand{\arcmm}{\ensuremath{\mathrm{^\prime}}}

\newcommand{\dotarcs}{\,\rlap{\hbox{$\mathrm{^\prime\hskip-0.1em^\prime}$}}{\hbox{$.$}}\,}

\newcommand{\dotsec}{\,\rlap{\hbox{$\mathrm{^s}$}}{\hbox{$.$}}\,}

\begin{document}
   \title{Cluster Abell 520: a perspective based on member galaxies}
   \subtitle{A cluster forming at the crossing of three filaments?}

   \author{M. Girardi\inst{1,2}
          \and
R. Barrena\inst{3}
          \and
           W. Boschin\inst{1,4}
\and
E. Ellingson\inst{5}}

   \offprints{M. Girardi, \email{girardi@oats.inaf.it}}

   \institute{Dipartimento di Astronomia of the Universit\`a degli
	      Studi di Trieste, via Tiepolo 11, I-34143 Trieste,
	      Italy\\ 
\and 
INAF - Osservatorio Astronomico di Trieste, via Tiepolo 11, I-34143
	      Trieste, Italy\\ 
\and
Instituto de Astrof\'{\i}sica de Canarias,
	      C/V\'{\i}a L\'actea s/n, E-38205 La Laguna (Tenerife),
	      Canary Islands, Spain\\
\and 
	      Fundaci\'on Galileo Galilei - INAF, Rambla Jos\'e Ana 
              Fern\'andez Perez 7, E-38712 Bre\~na Baja (La Palma), 
              Canary Islands, Spain\\
\and
              Center for Astrophysics and Space Astronomy, 389 UCB, 
              University of Colorado, Boulder, CO 80309\\
             }

   \date{Received  / Accepted }

\abstract {The connection of cluster mergers with the presence of
  extended, diffuse radio sources in galaxy clusters is still debated.
  An interesting case is the rich, merging cluster Abell 520,
  containing a radio halo. A recent gravitational analysis has shown
  in this cluster the presence of a massive dark core suggested to be
  a possible problem for the current cold dark matter paradigm.}  {We
  aim to obtain new insights into the internal dynamics of Abell 520
  analyzing velocities and positions of member galaxies.}{Our analysis
  is based on redshift data for 293 galaxies in the cluster field
  obtained combining new redshift data for 86 galaxies acquired at the
  TNG with data obtained by CNOC team and other few data from the
  literature.  We also use new photometric data obtained at the INT
  telescope. We combine galaxy velocities and positions to select 167
  cluster members around $z\sim0.201$.  We analyze the cluster
  structure using the weighted gap analysis, the KMM method, the
  Dressler-Shectman statistics and the analysis of the velocity
  dispersion profiles. We compare our results with those from X-ray,
  radio and gravitational lensing analyses.}{We compute a global
  line--of--sight (LOS) velocity dispersion of galaxies, $\sigma_{\rm
    v}=1066_{-61}^{+67}$ \kss.  We detect the presence of a high
  velocity group (HVG) with a rest--frame relative LOS velocity of
  ${\rm v_{\rm rf}}\sim 2000$ \ks with respect to the main system
  (MS). Using two alternative cluster models we estimate a mass range
  $M(<1 \hhh)=(4.0-9.6)$\mquaa.  We also find that the MS shows
  evidence of subclumps along two preferred directions.  The main,
  complex structure ${\cal NE}1$+${\cal NE}2$ (with a velocity
  comparable to that of the MS) and the ${\cal SW}$ structure (at
  ${\rm v_{\rm rf}}\sim +1100$ \kss) define the NE--SW direction, the
  same of the merger suggested by X--ray and radio data.  The ${\cal
    E}$ and ${\cal W}$ structures (at ${\rm v_{\rm rf}}\sim -1150$ and
  ${\rm v_{\rm rf}}\sim -300$ \kss) define the E--W direction.
  Moreover, we find no dynamical trace of an important structure
  around the lensing dark core. Rather, the HVG and a minor MS group,
  having different velocities, are roughly centered in the same
  position of the lensing dark core, i.e. are somewhat aligned with
  the LOS.}{ We find that Abell 520 is definitely a very complex
  system.  Our results suggest that we are looking at a cluster
  forming at the crossing of three filaments of the large scale
  structure. The filament aligned with the LOS and projected onto the
  center of the forming cluster might explain the apparent massive
  dark core shown by gravitational lensing analysis.}


   \maketitle
%

\section{Introduction}
\label{intr}

Clusters of galaxies are by now recognized to be not simple relaxed
structures, but rather they are evolving via merging processes in a
hierarchical fashion from poor groups to rich clusters. Much progress
has been made in recent years in the observations of the signatures of
merging processes (see Feretti et al. \cite{fer02b} for a general
review). A recent aspect of these investigations is the possible
connection of cluster mergers with the presence of extended, diffuse
radio sources: halos and relics. Cluster mergers have been suggested
to provide the large amount of energy necessary for electron
reacceleration and magnetic field amplification (Feretti \cite{fer99};
Feretti \cite{fer02a}; Sarazin \cite{sar02}). However, the question is
still debated since the diffuse radio sources are quite uncommon and
only recently we can study these phenomena on the basis of a
sufficient statistics (few dozen clusters up to $z\sim 0.3$, e.g.,
Giovannini et al. \cite{gio99}; see also Giovannini \& Feretti
\cite{gio02}; Feretti \cite{fer05}).

Growing evidence of the connection between diffuse radio emission and
cluster merging is based on X--ray data (e.g., B\"ohringer \&
Schuecker \cite{boh02}; Buote \cite{buo02}). Studies based on a large
number of clusters have found a significant relation between the radio
and the X--ray surface brightness (Govoni et al. \cite{gov01a},
\cite{gov01b}) and connections between the presence of
radio--halos/relics and irregular and bimodal X--ray surface
brightness distribution (Schuecker et al. \cite{sch01}).

Optical data are a powerful way to investigate the presence and the
dynamics of cluster mergers (e.g., Girardi \& Biviano \cite{gir02}),
too.  The spatial and kinematical analysis of member galaxies allow us
to detect and measure the amount of substructure, to identify and
analyze possible pre--merging clumps or merger remnants.  This optical
information is really complementary to X--ray information since
galaxies and intra--cluster medium react on different time scales
during a merger (see, e.g., numerical simulations by Roettiger et
al. \cite{roe97}). In this context we are conducting an intensive
observational and data analysis program to study the internal dynamics
of radio clusters by using member galaxies. Our program concerns both
massive clusters, where diffuse radio emissions are more frequently
found (e.g., Barrena et al. \cite{bar07b} and refs. therein), and
low--mass galaxy systems (Boschin et al. \cite{bos08}\footnote{please
  visit the web site of the DARC (Dynamical Analysis of Radio
  Clusters) project: http://adlibitum.oat.ts.astro.it/girardi/darc.}).

\begin{figure*}
\centering 
\includegraphics[width=16cm]{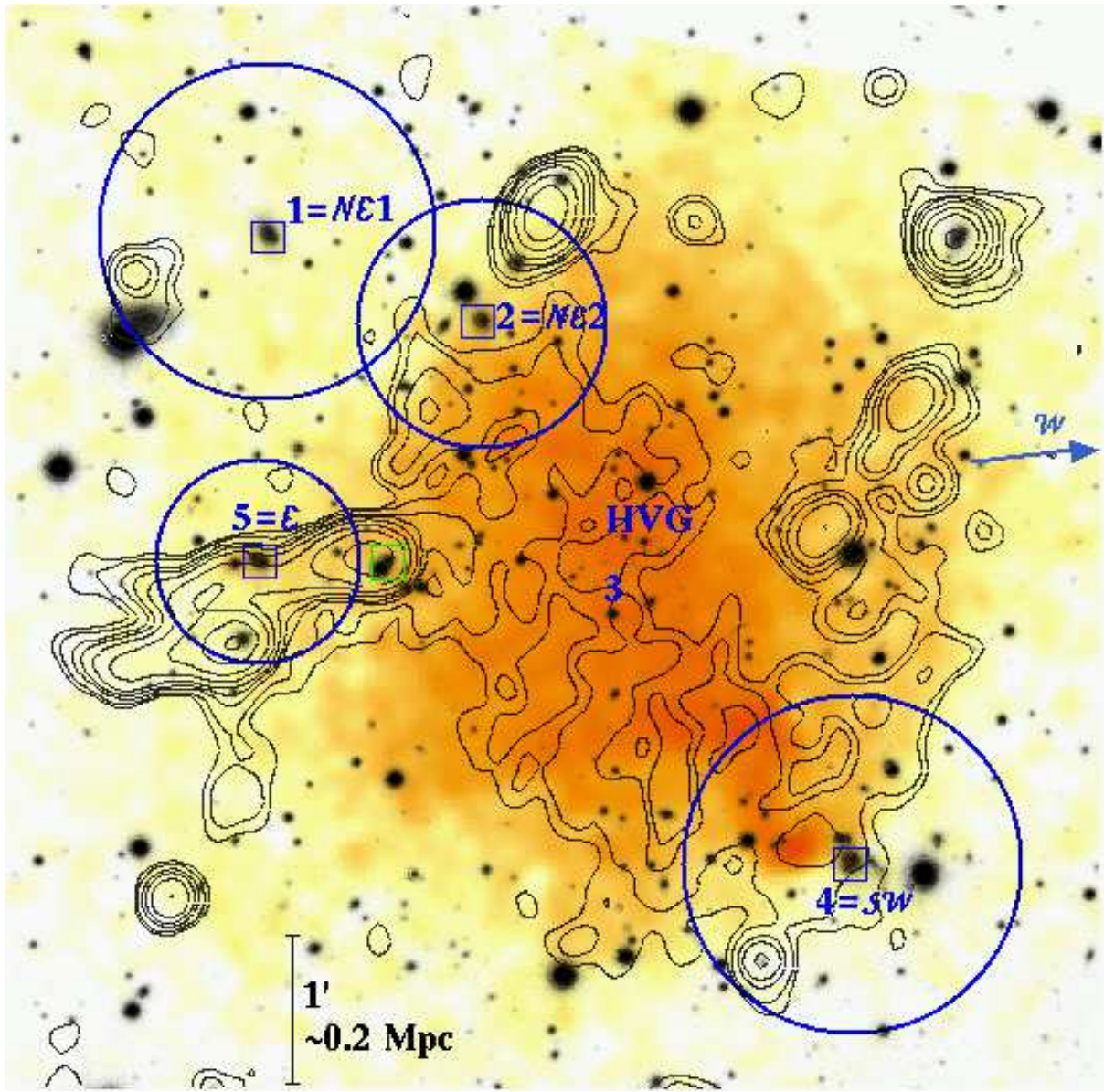}
\caption{ Multiwavelength picture of A520 (North is at the top and
  East to the left). A smoothed Chandra 0.5--2 keV image (orange and
  yellow colors) of the central region of A520 (courtesy of
  M. Markevitch - Markevitch et al. \cite{mar05}, X--ray point sources
  are removed) is superimposed to a $r^\prime$--band image taken with
  the WFC camera of the INT. The contour levels of a VLA radio image
  at 1.4 GHz (courtesy of F. Govoni - Govoni et al. \cite{gov01b}) are
  shown, too.  Main structures recovered by our analysis are
  highlighted (see Sect.~\ref{dyn} for more details).  Label HVG
  indicates the center of the high velocity group having a relative
  LOS velocity of ${\rm v_{\rm rf}} \sim 2000$ \ks with respect to the
  main system (MS).  Blue circles and numbers highlight the positions
  of the five peaks in the lensing mass distribution found by M07. The
  size of the circles indicate the regions where we find evidence for
  an individual, dynamically important structure of the MS. The name
  of each structure is indicated by the label close the corresponding
  M07 peak number and the blue small square indicates the central,
  luminous galaxy (i.e. galaxies IDs 204, 170, 106 and 205 for ${\cal
    NE}1$, ${\cal NE}2$, ${\cal SW}$ and ${\cal E}$, respectively).
  Finally, the green square indicates a head--tail radiogalaxy.  }
\label{figradio}
\end{figure*}

During our observational program we have conducted an intensive study
of the massive cluster Abell 520 (hereafter A520). This cluster shows
a radio halo, discovered by Giovannini et al. (\cite{gio99}), having a
low surface brightness with a clumpy structure  slightly elongated in
the NE--SW direction (Govoni et al. \cite{gov01b}; \cite{gov04}, see
Fig.~\ref{figradio}). 
 
A520, also known as MS 0451+02 in the EMSS catalog (Gioia et
al. \cite{gio90}), is a fairly rich, X--ray luminous, and hot cluster,
with a galaxy population characterized by a high velocity dispersion:
Abell richness class $=1$ (Abell et al. \cite{abe89}),
$L_\mathrm{X}$(0.1--2.4 keV)=14.20$\times 10^{44} \ h_{50}^{-2}$
erg\ s$^{-1}$ (Ebeling et al.~\cite{ebe96}); $T_\mathrm{X}= 7.1\pm0.7$
keV (Chandra data, Govoni et al. \cite{gov04}); $\sigma_{\rm
  v}=(988\pm 76)$ \ks (Carlberg et al. \cite{car96}).

First hints about the young dynamical status of this cluster came from
both X--ray and optical data (Le Fevre et al. \cite{lef94}; Gioia \&
Luppino \cite{gio94} and refs. therein). The complexity of its
structure was confirmed by analyses of ROSAT and Chandra X--ray data
(Govoni et al. \cite{gov01b}; \cite{gov04}). In particular, new
unprecedent insights were recovered from deep Chandra observations by
Markevitch et al. (\cite{mar05}). They revealed a prominent bow shock
indicating a cluster merger where a SW irregular structure consists of
dense, cool pieces of a cluster core that has been broken up by ram
pressure as it flew in from the NE direction (see
Fig.~\ref{figradio}).  The overall structure of the radio halo seems
connected with the cluster merger and may even suggest two distinct
components, a mushroom with a stem and a cap, where the main stem
component goes across the cluster along the NE--SW direction and the
cap ends at the bow shock (Govoni et al. \cite{gov01b}; Markevitch et
al. \cite{mar05}). 

The complex structure of A520 was also confirmed by gravitational
lensing analysis of Dahle et al. (\cite{dah02}), Mahdavi et
al. (\cite{mah07}, hereafter M07) and Okabe \& Umetsu
(\cite{oka08}). Okabe \& Umetsu (\cite{oka08}, based on Subaru data)
found a general good agreement between mass and galaxy luminosity
distribution. However, the detailed study of M07 based on the same
Subaru data and additional CFHT data pointed out a less clear
situation. M07 found four very significant peaks in the lensing mass
distribution. Among these, peaks No. 1, 2 and 4 correspond to peaks in
the galaxy distribution and give usual values for the mass--to--light
ratio.  Peak No. 3 corresponds to the central X-ray emission peak, but
is largely devoid of galaxies. This peak is characterized by a very
large mass--to--light value; thus to be referred as a ``massive dark 
core''. A region characterized by a somewhat low mass--to--light ratio
exists, too (less significant peak No. 5). This displacement between
galaxy and mass (i.e. dark matter, for the most part) remains very
puzzling. In fact, galaxies and cold dark matter (CDM), being both treated as
collisionless components, are expected to have similar behavior during
a cluster merger.  If confirmed by better observations, this situation
would be difficult to explain within the widely accepted 
CDM paradigm of cosmological structure formation (see M07 for
further discussions).

As for the analysis of the internal dynamics based on member galaxies,
Proust et al. (\cite{pro00}) found some evidence of substructure using
a sample of 21 galaxies, while the large data sample constructed by
the Canadian Network for Observational Cosmology (hereafter CNOC) team
(Carlberg et al.  \cite{car96}; Yee et al. \cite{yee96}) is still not
exploited a part from few individual galaxies in M07.  Recently, we
have carried out spectroscopic observations at the TNG telescope
giving new redshift data for 86 galaxies in the field of A520, as well
as photometric observations at the INT telescope. Our present analysis
is based on these optical data as well as on the large data sample
obtained by CNOC.

This paper is organized as follows.  We present our new optical data
in Sect.~2 and the complete redshift catalog with the addition of CNOC
and a few other data in Sect.~3. We present our results about 
global properties and substructure in Sect.~4.  We furtherly analyze
and discuss the dynamical status of A520 in Sect.~5. We draw our
conclusions in Sect.~6.

Unless otherwise stated, we give errors at the 68\% confidence level
(hereafter c.l.). Throughout this paper, we use $H_0=70\ h_{70}$  km s$^{-1}$
Mpc$^{-1}$ in a flat cosmology with $\Omega_0=0.3$ and
$\Omega_{\Lambda}=0.7$. In the adopted cosmology, 1\arcm corresponds
to $\sim 199$ \kpc at the cluster redshift.

\begin{figure*}
\centering 
\includegraphics[width=12cm]{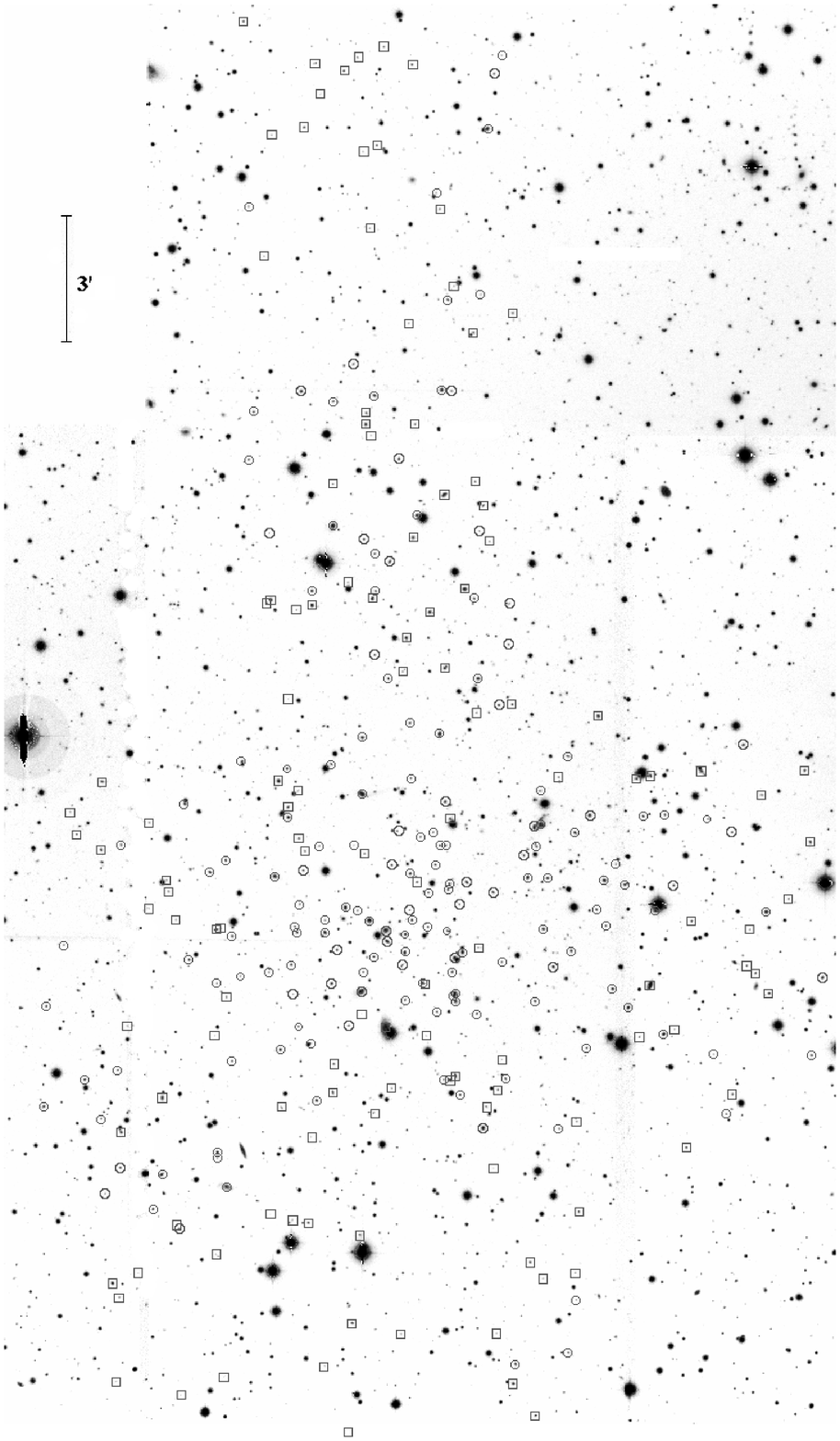}
\caption{INT $r^\prime$--band image of A520 (West at the top and North to the
left). Circles and boxes indicate cluster members and
non--member galaxies, respectively (see Table~\ref{catalogA520}).}
\label{figottico}
\end{figure*}

\section{New optical data}
\label{newd}

\subsection{Spectroscopic data}
\label{spec}

Multi--object spectroscopic observations of A520 were carried out at
the TNG telescope in December 2006. We used DOLORES/MOS with the LR--B
Grism 1, yielding a dispersion of 187 \AA/mm, and the Loral CCD of
$2048\times2048$ pixels (pixel size of 15 $\mu$m). This combination of
grating and detector results in a dispersion of 2.8 \AA/pix. We
observed three MOS masks for a total of 102 slits. We acquired three
exposures of 1800 s for each mask. Wavelength calibration was
performed using Helium--Argon lamps. Reduction of spectroscopic data
was carried out with the IRAF
\footnote{IRAF is distributed by the National Optical Astronomy
Observatories, which are operated by the Association of Universities
for Research in Astronomy, Inc., under cooperative agreement with the
National Science Foundation.} package.

Radial velocities were determined using the cross--correlation
technique (Tonry \& Davis \cite{ton79}) implemented in the RVSAO
package (developed at the Smithsonian Astrophysical Observatory
Telescope Data Center).  Each spectrum was correlated against six
templates for a variety of galaxy spectral types: E, S0, Sa, Sb, Sc,
Ir (Kennicutt \cite{ken92}).  The template producing the highest value
of $\cal R$, i.e., the parameter given by RVSAO and related to the
signal--to--noise of the correlation peak, was chosen.  Moreover, all
spectra and their best correlation functions were examined visually to
verify the redshift determination. The median value of $\cal R$ of our
successfully measured galaxy redshifts is $\sim 8$.  In nine cases (IDs
82, 86, 87, 143, 147 (QSO), 203, 229, 242 and 252; see Table
\ref{catalogA520}) we took the EMSAO redshift as a reliable estimate
of the redshift.  Our spectroscopic survey in the field of A520
consists of 86 spectra with a median nominal error on $cz$ of 60 \kss.
The nominal errors as given by the cross--correlation are known to be
smaller than the true errors (e.g., Malumuth et al.  \cite{mal92};
Bardelli et al. \cite{bar94}; Ellingson \& Yee \cite{ell94}; Quintana
et al. \cite{qui00}).  Double redshift determinations for the same
galaxy allowed us to estimate real intrinsic errors in data of the
same quality taken with the same instrument (Barrena et
al. \cite{bar07a}, \cite{bar07b}). Here we applied a similar
correction to our nominal errors, i.e.  hereafter we assume that true
errors are larger than nominal cross--correlation errors by a factor
1.5. Thus the median error on $cz$ is 90 \kss.

\subsection{Photometric data}
\label{phot}

As far as photometry is concerned, our observations were carried out
with the Wide Field Camera (WFC), mounted at the prime focus of the
2.5m INT telescope (located at Roque de los Muchachos observatory, La
Palma, Spain). We observed A520 in January 2008 in photometric
conditions and with a seeing of about 2 arcsec.

The WFC consists of a 4 chips mosaic covering a 30$\times$30 arcmin
field of view, with only a 20\% marginally vignetted area. We took 15
exposures of 360 s using the $r$--SDSS ($r^\prime$) filter, completing
a total of 5400 s in this band. Moreover, we developed a dithering
pattern in order to build a master ``supersky'' image that was used to
correct our images for fringing patterns (Gullixson \cite{gul92}). In
addition, the dithering helped us to clean cosmic rays and avoid gaps
between CCD chips. The complete reduction process (including flat
fielding, bias subtraction and bad columns elimination) yielded a
final co--added image where the variation of the sky was lower than
1\% in the whole frame.

Another effect associated with the wide field frames is the distortion
of the field. In order to match the photometric and spectroscopic
samples, a good astrometric solution taking into account these
distortions is needed. Using IRAF tasks and taking as reference the
USNO B1.0 catalog we were able to find an accurate astrometric
solution (rms$\sim$0.3 arcsec) across the full frame. The photometric
calibration was performed using Landolt standard fields with well
known $r^\prime$ magnitude. These fields were achieved during the
observation. We finally identified galaxies in our image and measured
their magnitudes with the SExtractor package (Bertin \& Arnouts
\cite{ber96}) and AUTOMAG procedure. In few cases (e.g., close
companion galaxies, galaxies close to defects of CCD), the standard
SExtractor photometric procedure failed. In these cases we computed
magnitudes by hand. This method consists in assuming a galaxy profile
of a typical elliptical and scale it to the maximum observed
value. The integration of this profile give us an estimate of the
magnitude. The idea of this method is similar to the PSF photometry,
but assuming a galaxy profile, more appropriate in this case.

As a final step, we estimated and corrected the galactic extinction,
$A_{r^\prime}$=0.12, from Burstein \& Heiles (\cite{bur82}) reddening
maps. We estimated that our photometric sample is complete down to
$r^\prime$ = 22.0 (23.0) for S/N = 5 (3) within the observed field.

\section{Construction of the galaxy catalog}
\label{galc}

In addition to our TNG data we considered redshifts coming from the
CNOC survey (Carlberg et al. \cite{car96}; Yee et al. \cite{yee96}).
A detailed description of the data reduction techniques for the
spectroscopic data is given in Yee et al. (\cite{yee96}). We
considered the 215 galaxies having a redshift determined via a
correlation significance parameter $\cal R$'$\gtrsim 3$ as suggested
by Yee et al. (\cite{yee96}), see also Ellingson \& Yee (\cite{ell94})
for the description of this parameter.  As for TNG data, we applied
the above correction to nominal errors leading to a median error on
$cz$ of $\sim 120$ \kss. This error is quite in agreement with the
error analysis performed by the CNOC authors (Ellingson \& Yee
\cite{ell94}).

Before to proceed with the merging between TNG and CNOC catalogs we
payed particular attention to their compatibility. Twelve galaxies in
the CNOC catalog are in common with our TNG catalog. Of these, one
(galaxy ID 215) can be considered as strongly discrepant with a
$\Delta cz$ difference of $\sim 1000$ \ks [$cz_{\rm TNG}=(58982\pm100)$
\ks vs. $cz_{\rm CNOC}=(59943\pm81)$ \kss]. For this galaxy a redshift by
Newberry et al. (\cite{new88}) also exists and it is in agreement with
the TNG redshift. For the remaining eleven galaxies we compared the
TNG and CNOC determinations computing the mean and the rms of the
variable $(z_1-z_2)/\sqrt{err_1^2+err_2^2}$, where $z_1$ comes from
TNG, and $z_2$ from CNOC. We obtained $mean=0.53\pm0.66$ and
$rms=2.2$, to be compared with the expected values of 0 and 1. The
resulting mean shows that the two sets of measurements are consistent
with having the same velocity zero--point according to the
$\chi^2$--test. The high value of rms suggests that the errors are
still underestimated. However, when rejecting another two slightly
discrepant determinations ($\Delta cz \sim 700$ \ks for IDs 253 and
72) we obtained $mean=0.55\pm0.46$ and $rms=1.4$, in good agreement
with the expected values of 0 and 1. We decided to take our TNG
redshifts for the galaxies IDs 215, 253 and 72 and combine TNG and
CNOC data using the weighted mean of the two redshift determinations
and the corresponding errors for the remaining nine common
galaxies. In total, we added another new 203 galaxies from CNOC
obtaining a merged catalog of 289 galaxies.

Finally, we considered the catalog of galaxies in the field of A520
published by Proust et al. (\cite{pro00}, their Table~1). For three
galaxies (the 2nd, 16th, and 19th) Proust et al. list only redshifts
coming from old previous literature data. These galaxies are already
present in our TNG catalog and we verified the agreement between our
and previous redshift values. Out of the 24 galaxies measured by Proust
et al., we considered only the 13 galaxies with $\cal R \gtrsim$ 3 and
one galaxy with redshift measured on the emission line
H$\alpha$. After having applied the correction to their nominal
redshift, we checked the compatibility with our TNG+CNOC catalog using
the method described above. We found nine galaxies in common with our
catalog for which we obtain $mean=-0.23\pm0.46$ and $rms=1.4$, in 
agreement with the expected values of 0 and 1. We combined TNG+CNOC
catalog and Proust et al.  data using the weighted mean of the two
redshift determinations and the corresponding error for the nine
galaxies in common.  We added another new four galaxies by Proust et
al., two of which are very bright galaxies.


\begin{table}[!ht]
        \caption[]{Velocity catalog of 293 spectroscopically measured
galaxies in the field of A520. In Col.~1, IDs in italics indicate
non--cluster galaxies.}
         \label{catalogA520}
              $$ 
           \begin{array}{r c c r r c}
            \hline
            \noalign{\smallskip}
            \hline
            \noalign{\smallskip}

\mathrm{ID} & \mathrm{\alpha},\mathrm{\delta}\,(\mathrm{J}2000)  & r^\prime & \mathrm{v}\,\,\,\,\,\,\, & \mathrm{\Delta}\mathrm{v} & \mathrm{Source}\\
  & & &\mathrm{(\,km}&\mathrm{s^{-1}\,)}& \\
            \hline
            \noalign{\smallskip}  

\textit{1}    &04\ 52\ 48.06 ,+03\ 00\ 43.2&       19.62&  94953 &  99& \mathrm{C}\\     
\textit{2}    &04\ 52\ 50.27 ,+02\ 57\ 24.7&       20.13&  68383 & 153& \mathrm{C}\\     
3             &04\ 52\ 50.93 ,+02\ 54\ 37.9&       20.87&  60186 & 140& \mathrm{C}\\     
\textit{4}    &04\ 52\ 51.30 ,+02\ 58\ 01.7&       20.21&  98922 & 104& \mathrm{C}\\     
\textit{5}    &04\ 52\ 51.91 ,+02\ 56\ 44.9&       19.50&  68173 & 112& \mathrm{C}\\     
\textit{6}    &04\ 52\ 51.94 ,+02\ 59\ 04.0&       19.75& 108794 & 225& \mathrm{C}\\     
\textit{7}    &04\ 52\ 52.60 ,+02\ 58\ 21.2&       19.68&  56568 &  90& \mathrm{C}\\     
8             &04\ 52\ 52.68 ,+02\ 54\ 48.6&       19.89&  60738 & 117& \mathrm{C}\\     
\textit{9}    &04\ 52\ 54.83 ,+02\ 58\ 54.2&       21.32& 103989 & 225& \mathrm{C}\\     
10            &04\ 52\ 57.91 ,+02\ 54\ 57.3&       18.25&  60918 & 108& \mathrm{C}\\     
\textit{11}   &04\ 52\ 58.01 ,+02\ 59\ 16.8&       19.66&  54316 & 126& \mathrm{C}\\     
\textit{12}   &04\ 52\ 58.76 ,+03\ 00\ 02.7&       20.83&  90780 &  94& \mathrm{C}\\     
\textit{13}   &04\ 52\ 59.62 ,+02\ 57\ 34.1&       19.89&  88858 & 112& \mathrm{C}\\     
\textit{14}   &04\ 53\ 00.24 ,+02\ 57\ 53.3&       20.92&  99564 &  90& \mathrm{C}\\     
15            &04\ 53\ 04.06 ,+02\ 56\ 09.4&       20.61&  59293 & 130& \mathrm{C}\\     
\textit{16}   &04\ 53\ 05.59 ,+02\ 56\ 04.1&       19.99&  64479 & 126& \mathrm{C}\\     
17            &04\ 53\ 05.61 ,+03\ 00\ 35.1&       20.16&  60606 &  99& \mathrm{C}\\     
\textit{18}   &04\ 53\ 07.50 ,+02\ 57\ 43.4&       20.61& 116442 & 225& \mathrm{C}\\     
\textit{19}   &04\ 53\ 10.30 ,+03\ 00\ 12.7&       20.98&  91469 & 135& \mathrm{C}\\     
\textit{20}   &04\ 53\ 12.90 ,+02\ 55\ 43.5&       20.84&  77805 & 135& \mathrm{C}\\     
21            &04\ 53\ 13.68 ,+02\ 55\ 05.3&       20.58&  59383 & 117& \mathrm{C}\\     
22            &04\ 53\ 14.19 ,+02\ 55\ 52.0&       19.39&  61964 & 108& \mathrm{C}\\     
\textit{23}   &04\ 53\ 15.37 ,+02\ 54\ 19.6&       19.90&  64440 & 162& \mathrm{C}\\   
\textit{24}   &04\ 53\ 16.47 ,+02\ 56\ 47.3&       20.89&  98509 & 126& \mathrm{C}\\   
\textit{25}   &04\ 53\ 17.28 ,+02\ 55\ 15.1&       19.72&  77751 & 112& \mathrm{C}\\     
26            &04\ 53\ 20.43 ,+02\ 58\ 05.0&       19.90&  60741 & 130& \mathrm{C}\\   
27            &04\ 53\ 22.74 ,+02\ 55\ 59.6&       18.25&  61547 & 108& \mathrm{C}\\   
28            &04\ 53\ 22.83 ,+02\ 55\ 45.4&       19.77&  62471 &  99& \mathrm{C}\\     
29            &04\ 53\ 23.03 ,+02\ 59\ 18.6&       18.62&  59722 & 104& \mathrm{C}\\     
30            &04\ 53\ 23.35 ,+02\ 57\ 34.8&       18.79&  62225 & 112& \mathrm{C}\\     
31            &04\ 53\ 23.95 ,+02\ 58\ 33.0&       19.74&  60495 & 130& \mathrm{C}\\   
\textit{32}   &04\ 53\ 24.99 ,+02\ 57\ 46.2&       19.64&  78258 & 126& \mathrm{C}\\     
33            &04\ 53\ 25.04 ,+03\ 00\ 25.7&       19.48&  60057 & 126& \mathrm{C}\\   
\textit{34}   &04\ 53\ 26.03 ,+02\ 56\ 36.7&       20.21&  98524 & 130& \mathrm{C}\\     
\textit{35}   &04\ 53\ 26.11 ,+02\ 57\ 46.7&       18.70&  99015 & 122& \mathrm{C}\\   
       
               \noalign{\smallskip}			    
            \hline					    
            \noalign{\smallskip}			    
            \hline					    
         \end{array}
     $$ 
         \end{table}
\addtocounter{table}{-1}
\begin{table}[!ht]
          \caption[ ]{Continued.}
     $$ 
           \begin{array}{r c c r r c}
            \hline
            \noalign{\smallskip}
            \hline
            \noalign{\smallskip}

\mathrm{ID} & \mathrm{\alpha},\mathrm{\delta}\,(\mathrm{J}2000)  & r^\prime & \mathrm{v}\,\,\,\,\,\,\, & \mathrm{\Delta}\mathrm{v}& \mathrm{Source} \\
  & & &\mathrm{(\,km}&\mathrm{s^{-1}\,)}& \\
            \hline
            \noalign{\smallskip}

\textit{36}   &04\ 53\ 27.14 ,+02\ 57\ 38.8&       21.22& 139946 & 225& \mathrm{C}\\     
37            &04\ 53\ 29.33 ,+02\ 56\ 58.9&       19.22&  60729 &  99& \mathrm{C}\\     
38            &04\ 53\ 29.61 ,+03\ 00\ 31.8&       19.66&  58588 &  86& \mathrm{C}\\     
\textit{39}   &04\ 53\ 31.36 ,+02\ 55\ 09.9&       20.08&  79496 & 130& \mathrm{C}\\     
\textit{40}   &04\ 53\ 31.79 ,+02\ 58\ 32.1&       20.76&  69786 & 126& \mathrm{C}\\     
\textit{41}   &04\ 53\ 32.65 ,+02\ 55\ 53.7&       18.14&  36736 & 112& \mathrm{C}\\     
\textit{42}   &04\ 53\ 33.63 ,+02\ 54\ 57.9&       19.61&  63283 & 140& \mathrm{C}\\     
43            &04\ 53\ 34.62 ,+02\ 56\ 32.4&       19.58&  59880 &  99& \mathrm{C}\\     
44            &04\ 53\ 35.76 ,+02\ 58\ 31.5&       17.29&  59488 & 122& \mathrm{C}\\     
45            &04\ 53\ 36.05 ,+02\ 55\ 03.4&       20.86&  59353 & 126& \mathrm{C}\\     
46            &04\ 53\ 36.54 ,+03\ 00\ 01.6&       21.03&  59401 & 144& \mathrm{C}\\     
\textit{47}   &04\ 53\ 36.76 ,+02\ 56\ 37.3&       18.85&  64659 & 117& \mathrm{C}\\     
48            &04\ 53\ 36.99 ,+02\ 57\ 47.3&       19.84&  59320 &  99& \mathrm{C}\\     
\textit{49}   &04\ 53\ 37.01 ,+02\ 54\ 48.7&       20.90&  79073 & 104& \mathrm{C}\\     
50            &04\ 53\ 38.36 ,+02\ 57\ 31.6&       19.88&  59272 & 130& \mathrm{C}\\     
51            &04\ 53\ 39.06 ,+02\ 57\ 10.3&       18.99&  59707 & 112& \mathrm{C}\\     
\textit{52}   &04\ 53\ 41.08 ,+02\ 58\ 09.1&       20.85&  62576 &  94& \mathrm{C}\\     
\textit{53}   &04\ 53\ 41.56 ,+02\ 55\ 23.4&       18.18&  62962 & 108& \mathrm{C}\\     
54            &04\ 53\ 41.88 ,+02\ 57\ 29.9&       19.84&  59832 & 135& \mathrm{C}\\     
55            &04\ 53\ 41.97 ,+02\ 59\ 00.3&       19.26&  60168 & 130& \mathrm{C}\\     
56            &04\ 53\ 42.44 ,+02\ 55\ 09.4&       19.68&  58241 & 126& \mathrm{C}\\     
\textit{57}   &04\ 53\ 42.56 ,+02\ 57\ 33.9&       18.73&  64665 & 122& \mathrm{C}\\     
\textit{58}   &04\ 53\ 42.86 ,+02\ 59\ 58.9&       18.55&  44786 & 126& \mathrm{C}\\   
59            &04\ 53\ 42.89 ,+02\ 54\ 19.3&       20.37&  58729 & 117& \mathrm{C}\\   
\textit{60}   &04\ 53\ 43.19 ,+03\ 00\ 04.2&       20.18&  98703 & 158& \mathrm{C}\\     
\textit{61}   &04\ 53\ 43.32 ,+02\ 58\ 59.9&       18.67&  65472 &  94& \mathrm{C}\\   
\textit{62}   &04\ 53\ 43.78 ,+02\ 59\ 22.2&       20.96&  79478 &  94& \mathrm{C}\\   
\textit{63}   &04\ 53\ 43.81 ,+02\ 56\ 12.9&       18.65&  65247 & 130& \mathrm{C}\\     
\textit{64}   &04\ 53\ 46.26 ,+02\ 56\ 45.8&       18.68&  64569 & 108& \mathrm{C}\\     
65            &04\ 53\ 46.77 ,+02\ 54\ 19.9&       20.07&  61199 & 158& \mathrm{C}\\     
66            &04\ 53\ 47.93 ,+02\ 57\ 30.2&       20.05&  59578 & 117& \mathrm{C}\\   
\textit{67}   &04\ 53\ 49.09 ,+02\ 55\ 50.5&       18.97&  75509 & 108& \mathrm{C}\\     
\textit{68}   &04\ 53\ 49.45 ,+02\ 56\ 50.2&       20.27&  65478 & 126& \mathrm{C}\\   
69            &04\ 53\ 50.05 ,+02\ 55\ 02.7&       18.64&  59467 & 135& \mathrm{C}\\     
70            &04\ 53\ 50.16 ,+02\ 57\ 10.2&       19.60&  61436 & 122& \mathrm{C}\\   

                \noalign{\smallskip}			    
            \hline					    
            \noalign{\smallskip}			    
            \hline					    
         \end{array}
     $$ 
         \end{table}

\addtocounter{table}{-1}
\begin{table}[!ht]
          \caption[ ]{Continued.}
     $$ 
           \begin{array}{r c c r r c}

            \hline
            \noalign{\smallskip}
            \hline
            \noalign{\smallskip}

\mathrm{ID} & \mathrm{\alpha},\mathrm{\delta}\,(\mathrm{J}2000)  & r^\prime & \mathrm{v}\,\,\,\,\,\,\, & \mathrm{\Delta}\mathrm{v}& \mathrm{Source}  \\
  & & &\mathrm{(\,km}&\mathrm{s^{-1}\,)}& \\
            \hline
            \noalign{\smallskip} 

\textit{71}   &04\ 53\ 52.22 ,+02\ 59\ 32.5&       21.78&  65364 & 117& \mathrm{C}   \\     
\textit{72}   &04\ 53\ 52.44 ,+02\ 54\ 14.7&       19.91&  65804 &  58& \mathrm{T}   \\     
73            &04\ 53\ 52.53 ,+02\ 54\ 32.7&       19.57&  62582 & 130& \mathrm{C}   \\     
\textit{74}   &04\ 53\ 53.31 ,+02\ 55\ 04.5&       20.59& 115810 & 225& \mathrm{C}   \\     
\textit{75}   &04\ 53\ 53.46 ,+02\ 52\ 12.1&       18.86&  67466 & 102& \mathrm{T}   \\     
76            &04\ 53\ 54.33 ,+02\ 56\ 39.2&       19.29&  61182 & 126& \mathrm{C}   \\     
77            &04\ 53\ 55.31 ,+02\ 55\ 56.5&       18.26&  60198 & 122& \mathrm{C}   \\     
78            &04\ 53\ 55.74 ,+02\ 57\ 46.6&       18.83&  59254 & 126& \mathrm{C}   \\     
79            &04\ 53\ 55.99 ,+02\ 48\ 45.7&       19.21&  61134 &  60& \mathrm{T}   \\     
80            &04\ 53\ 57.31 ,+02\ 52\ 54.4&       19.44&  62808 & 135& \mathrm{T}   \\     
81            &04\ 53\ 58.20 ,+03\ 00\ 37.8&       19.49&  59476 & 140& \mathrm{C}   \\     
\textit{82}   &04\ 53\ 58.33 ,+02\ 47\ 18.1&       19.47& 115868 &  64& \mathrm{T}   \\     
83            &04\ 53\ 58.35 ,+02\ 58\ 30.5&       19.54&  60492 & 748& \mathrm{C}   \\     
\textit{84}   &04\ 53\ 58.54 ,+02\ 49\ 45.0&       19.10&  65115 &  87& \mathrm{T}   \\     
85            &04\ 53\ 58.83 ,+02\ 59\ 33.4&       19.34&  60357 &  99& \mathrm{C}   \\     
\textit{86}   &04\ 53\ 59.04 ,+02\ 50\ 57.3&       19.67&  64947 &  27& \mathrm{T}   \\     
\textit{87}   &04\ 53\ 59.29 ,+02\ 53\ 07.2&       20.41&  76130 &  21& \mathrm{T}   \\     
\textit{88}   &04\ 53\ 59.36 ,+02\ 51\ 17.1&       19.45&  76190 &  63& \mathrm{T}   \\     
89            &04\ 53\ 59.59 ,+02\ 56\ 36.5&       20.04&  62429 & 148& \mathrm{C}   \\     
\textit{90}   &04\ 53\ 59.99 ,+02\ 59\ 45.1&       18.40&  49274 & 117& \mathrm{C}   \\     
\textit{91}   &04\ 54\ 00.34 ,+03\ 03\ 55.9&       19.39&  70008 &  80& \mathrm{T}   \\     
92            &04\ 54\ 00.56 ,+02\ 53\ 32.9&       19.58&  61319 & 144& \mathrm{C}   \\     
\textit{93}   &04\ 54\ 00.73 ,+02\ 48\ 19.2&       19.66&  64849 &  88& \mathrm{T}   \\   
\textit{94}   &04\ 54\ 00.88 ,+02\ 59\ 16.5&       20.62&  77280 & 135& \mathrm{C}   \\   
95            &04\ 54\ 01.15 ,+02\ 57\ 45.6&       17.35&  62154 &  51& \mathrm{C+P} \\     
96            &04\ 54\ 01.76 ,+02\ 55\ 47.4&       19.17&  58327 &  90& \mathrm{C}   \\   
97            &04\ 54\ 02.37 ,+03\ 01\ 58.2&       19.34&  59542 &  68& \mathrm{T}   \\   
\textit{98}   &04\ 54\ 02.49 ,+02\ 59\ 31.0&       19.25&  49364 & 320& \mathrm{C}   \\     
99            &04\ 54\ 02.71 ,+02\ 50\ 36.2&       19.61&  58860 &  68& \mathrm{T}   \\     
100           &04\ 54\ 02.88 ,+02\ 52\ 22.6&       18.63&  60818 &  76& \mathrm{T}   \\     
101           &04\ 54\ 02.90 ,+02\ 51\ 06.7&       19.61&  60461 & 116& \mathrm{T}   \\   
102           &04\ 54\ 03.05 ,+02\ 49\ 34.7&       20.24&  58347 & 117& \mathrm{T}   \\     
\textit{103}  &04\ 54\ 03.26 ,+03\ 04\ 40.5&       20.83& 152875 & 112& \mathrm{T}   \\   
\textit{104}  &04\ 54\ 03.36 ,+02\ 55\ 40.3&       19.74&  70034 & 144& \mathrm{C}   \\     
105           &04\ 54\ 03.45 ,+02\ 59\ 30.6&       18.79&  60516 & 122& \mathrm{C}   \\   

              \noalign{\smallskip}			    
            \hline					    
            \noalign{\smallskip}			    
            \hline					    
         \end{array}\\
     $$ 
         \end{table}

\addtocounter{table}{-1}
\begin{table}[!ht]
          \caption[ ]{Continued.}
     $$ 
           \begin{array}{r c c r r c}
            \hline
            \noalign{\smallskip}
            \hline
            \noalign{\smallskip}

\mathrm{ID} & \mathrm{\alpha},\mathrm{\delta}\,(\mathrm{J}2000)  & r^\prime & \mathrm{v}\,\,\,\,\,\,\, & \mathrm{\Delta}\mathrm{v}& \mathrm{Source} \\
 & & &\mathrm{(\,km}&\mathrm{s^{-1}\,)}& \\

            \hline
            \noalign{\smallskip} 

106           &04\ 54\ 03.82 ,+02\ 53\ 32.4&       17.10&  61277 & 114& \mathrm{T}     \\     
107           &04\ 54\ 03.96 ,+02\ 53\ 40.7&       17.98&  64646 &  70& \mathrm{T+C}   \\     
\textit{108}  &04\ 54\ 04.18 ,+03\ 02\ 48.4&       20.84&  55511 & 104& \mathrm{T}     \\     
109           &04\ 54\ 04.30 ,+02\ 49\ 00.8&       20.01&  59783 & 141& \mathrm{T}     \\     
110           &04\ 54\ 04.54 ,+02\ 52\ 43.4&       19.08&  60827 &  98& \mathrm{T}     \\     
111           &04\ 54\ 04.59 ,+02\ 56\ 54.2&       20.12&  61145 & 112& \mathrm{P}     \\     
112           &04\ 54\ 04.67 ,+02\ 56\ 04.0&       20.18&  60327 & 148& \mathrm{C}     \\     
\textit{113}  &04\ 54\ 05.13 ,+02\ 47\ 09.1&       19.26&  65020 &  80& \mathrm{T}     \\     
114           &04\ 54\ 05.14 ,+02\ 56\ 22.2&       20.02&  59536 & 130& \mathrm{C}     \\     
\textit{115}  &04\ 54\ 05.38 ,+03\ 04\ 31.0&       20.26&  70903 &  92& \mathrm{T}     \\     
\textit{116}  &04\ 54\ 05.44 ,+02\ 59\ 15.5&       19.76&  66554 & 117& \mathrm{C}     \\     
117           &04\ 54\ 05.92 ,+02\ 55\ 54.2&       19.92&  59916 &  72& \mathrm{C+P}   \\     
118           &04\ 54\ 05.92 ,+02\ 55\ 46.0&       20.15&  58669 & 108& \mathrm{C}     \\     
119           &04\ 54\ 05.93 ,+02\ 53\ 37.3&       19.82&  59458 & 153& \mathrm{C}     \\     
120           &04\ 54\ 06.02 ,+02\ 57\ 56.7&       21.22&  62839 & 104& \mathrm{C}     \\     
121           &04\ 54\ 06.12 ,+02\ 58\ 46.1&       20.20&  62018 & 450& \mathrm{C}     \\     
122           &04\ 54\ 06.32 ,+03\ 03\ 27.9&       19.67&  59639 & 128& \mathrm{T}     \\     
\textit{123}  &04\ 54\ 06.64 ,+02\ 59\ 06.2&       20.09&  98850 & 122& \mathrm{C}     \\     
124           &04\ 54\ 06.77 ,+02\ 53\ 55.4&       19.59&  61473 &  71& \mathrm{T+C}   \\     
\textit{125}  &04\ 54\ 06.78 ,+02\ 57\ 41.6&       20.39&  66990 &  54& \mathrm{C+P}   \\     
\textit{126}  &04\ 54\ 06.82 ,+03\ 03\ 56.0&       19.38&  70420 &  64& \mathrm{T}     \\     
127           &04\ 54\ 07.47 ,+02\ 51\ 44.6&       20.70&  59358 &  93& \mathrm{T}     \\     
128           &04\ 54\ 07.62 ,+03\ 00\ 59.2&       19.31&  59490 &  46& \mathrm{T+C}   \\   
129           &04\ 54\ 07.74 ,+02\ 56\ 00.4&       19.78&  60861 & 104& \mathrm{C}     \\   
130           &04\ 54\ 07.83 ,+02\ 57\ 02.7&       19.82&  61140 & 122& \mathrm{C}     \\     
131           &04\ 54\ 08.47 ,+02\ 59\ 08.4&       20.03&  60060 & 117& \mathrm{C}     \\   
132           &04\ 54\ 08.63 ,+02\ 56\ 36.1&       19.19&  62695 &  94& \mathrm{C}     \\   
133           &04\ 54\ 08.78 ,+03\ 01\ 21.4&       19.10&  60931 &  72& \mathrm{T}     \\     
134           &04\ 54\ 08.87 ,+02\ 53\ 49.5&       18.69&  57872 & 117& \mathrm{C}     \\     
135           &04\ 54\ 08.90 ,+02\ 53\ 21.5&       19.11&  59219 &  66& \mathrm{T}     \\     
136           &04\ 54\ 09.03 ,+02\ 52\ 00.8&       18.63&  63622 &  66& \mathrm{T}     \\   
137           &04\ 54\ 09.06 ,+02\ 59\ 48.8&       17.93&  60075 &  76& \mathrm{C}     \\     
138           &04\ 54\ 09.37 ,+02\ 55\ 15.8&       18.97&  61125 & 117& \mathrm{C}     \\   
139           &04\ 54\ 09.41 ,+02\ 50\ 22.7&       19.66&  60440 & 104& \mathrm{T}     \\     
140           &04\ 54\ 09.41 ,+02\ 51\ 32.1&       19.38&  61024 &  48& \mathrm{T}     \\

              \noalign{\smallskip}			    
            \hline					    
            \noalign{\smallskip}			    
            \hline					    
         \end{array}\\
     $$ 
\end{table}

\addtocounter{table}{-1}
\begin{table}[!ht]
          \caption[ ]{Continued.}
     $$ 
           \begin{array}{r c c r r c}

            \hline
            \noalign{\smallskip}
            \hline
            \noalign{\smallskip}

\mathrm{ID} & \mathrm{\alpha},\mathrm{\delta}\,(\mathrm{J}2000)  & r^\prime & \mathrm{v}\,\,\,\,\,\,\, & \mathrm{\Delta}\mathrm{v}& \mathrm{Source} \\
 & & &\mathrm{(\,km}&\mathrm{s^{-1}\,)}& \\
            \hline
            \noalign{\smallskip} 

\textit{141}  &04\ 54\ 09.42 ,+02\ 56\ 26.3&       19.89&  67036 &  94& \mathrm{C}  \\     
142           &04\ 54\ 09.55 ,+02\ 55\ 40.5&       19.20&  58879 & 112& \mathrm{C}  \\     
\textit{143}  &04\ 54\ 09.60 ,+03\ 02\ 22.9&       20.83& 181270 &  51& \mathrm{T}  \\     
144           &04\ 54\ 10.10 ,+02\ 55\ 42.2&       19.69&  60600 & 117& \mathrm{C}  \\     
145           &04\ 54\ 10.31 ,+02\ 54\ 38.9&       19.41&  60963 &  51& \mathrm{T+C}\\     
146           &04\ 54\ 10.41 ,+02\ 56\ 09.9&       20.02&  58280 & 148& \mathrm{C}  \\     
\textit{147}  &04\ 54\ 10.52 ,+02\ 47\ 39.2&       19.83& 704502 & 100& \mathrm{T}  \\     
\textit{148}  &04\ 54\ 10.69 ,+03\ 02\ 20.0&       20.11& 152100 &  96& \mathrm{T}  \\     
149           &04\ 54\ 11.48 ,+02\ 55\ 25.8&       20.26&  57941 & 108& \mathrm{C}  \\     
150           &04\ 54\ 11.69 ,+02\ 59\ 13.1&       19.98&  62291 & 135& \mathrm{C}  \\     
151           &04\ 54\ 11.79 ,+02\ 48\ 10.7&       18.28&  58729 &  56& \mathrm{T}  \\     
152           &04\ 54\ 11.80 ,+02\ 52\ 11.4&       19.60&  60376 &  54& \mathrm{T}  \\     
153           &04\ 54\ 11.82 ,+02\ 50\ 48.1&       19.11&  59577 & 182& \mathrm{T}  \\     
154           &04\ 54\ 11.93 ,+02\ 58\ 07.8&       18.23&  62292 &  47& \mathrm{C+P}\\     
155           &04\ 54\ 12.08 ,+02\ 56\ 36.5&       20.77&  60708 & 148& \mathrm{C}  \\     
156           &04\ 54\ 12.19 ,+02\ 57\ 50.7&       20.48&  58417 & 153& \mathrm{C}  \\     
\textit{157}  &04\ 54\ 12.31 ,+03\ 02\ 47.8&       20.89&  85271 & 188& \mathrm{T}  \\     
\textit{158}  &04\ 54\ 12.77 ,+02\ 49\ 56.5&       19.66&  82005 &  78& \mathrm{T}  \\     
159           &04\ 54\ 13.04 ,+02\ 56\ 33.2&       19.36&  60084 & 144& \mathrm{C}  \\     
160           &04\ 54\ 13.14 ,+02\ 57\ 33.8&       17.70&  60115 &  24& \mathrm{P}  \\     
161           &04\ 54\ 13.16 ,+02\ 58\ 36.6&       20.25&  62495 & 135& \mathrm{C}  \\     
\textit{162}  &04\ 54\ 13.34 ,+03\ 02\ 08.8&       20.65&  99114 & 106& \mathrm{T}  \\        
163           &04\ 54\ 13.35 ,+02\ 51\ 58.1&       20.05&  60776 &  96& \mathrm{T}  \\   
\textit{164}  &04\ 54\ 13.50 ,+02\ 48\ 33.7&       20.45&  75930 & 180& \mathrm{T}  \\   
165           &04\ 54\ 13.68 ,+02\ 56\ 10.2&       19.78&  59653 &  99& \mathrm{C}  \\     
166           &04\ 54\ 13.74 ,+02\ 53\ 26.7&       19.14&  60519 &  72& \mathrm{T}  \\      
167           &04\ 54\ 13.80 ,+02\ 59\ 19.4&       19.81&  59059 & 130& \mathrm{C}  \\   
168           &04\ 54\ 14.01 ,+02\ 55\ 42.5&       19.63&  59383 & 122& \mathrm{C}  \\     
\textit{169}  &04\ 54\ 14.09 ,+03\ 01\ 05.1&       20.39&  40586 & 144& \mathrm{C}  \\     
170           &04\ 54\ 14.10 ,+02\ 57\ 09.9&       17.29&  59506 &  69& \mathrm{P}  \\     
\textit{171}  &04\ 54\ 14.17 ,+03\ 01\ 10.3&       18.89&  66961 & 140& \mathrm{C}  \\   
172           &04\ 54\ 14.34 ,+02\ 58\ 36.5&       18.50&  60111 & 104& \mathrm{C}  \\     
173           &04\ 54\ 14.36 ,+02\ 59\ 16.3&       19.41&  58513 & 117& \mathrm{C}  \\   
174           &04\ 54\ 14.40 ,+02\ 56\ 42.2&       19.02&  60762 &  99& \mathrm{C}  \\     
175           &04\ 54\ 14.79 ,+03\ 00\ 49.0&       19.01&  58513 &  75& \mathrm{T+C}\\   

              \noalign{\smallskip}			    
            \hline					    
            \noalign{\smallskip}			    
            \hline					    
         \end{array}\\
     $$ 
         \end{table}

\addtocounter{table}{-1}
\begin{table}[!ht]
          \caption[ ]{Continued.}
     $$ 
           \begin{array}{r c c r r c}

            \hline
            \noalign{\smallskip}
            \hline
            \noalign{\smallskip}

\mathrm{ID} & \mathrm{\alpha},\mathrm{\delta}\,(\mathrm{J}2000)  & r^\prime & \mathrm{v}\,\,\,\,\,\,\, & \mathrm{\Delta}\mathrm{v}& \mathrm{Source} \\
 & & &\mathrm{(\,km}&\mathrm{s^{-1}\,)}& \\
            \hline
            \noalign{\smallskip} 

176           &04\ 54\ 15.09 ,+02\ 57\ 07.8&       18.06&  59163 &  80& \mathrm{C+P}\\     
\textit{177}  &04\ 54\ 15.55 ,+02\ 54\ 58.3&       20.26&  18338 & 270& \mathrm{C}  \\     
178           &04\ 54\ 15.76 ,+02\ 52\ 46.9&       19.09&  62064 & 122& \mathrm{T}  \\     
179           &04\ 54\ 15.89 ,+03\ 04\ 47.6&       20.62&  58757 & 122& \mathrm{T}  \\     
180           &04\ 54\ 15.95 ,+02\ 58\ 19.1&       19.43&  60267 & 135& \mathrm{C}  \\     
181           &04\ 54\ 16.01 ,+02\ 55\ 20.7&       18.33&  60954 & 104& \mathrm{C}  \\     
182           &04\ 54\ 16.06 ,+02\ 56\ 42.8&       18.64&  58821 &  94& \mathrm{C+P}\\     
183           &04\ 54\ 16.56 ,+02\ 57\ 26.7&       19.67&  58876 & 126& \mathrm{C}  \\     
184           &04\ 54\ 16.57 ,+02\ 55\ 31.8&       19.35&  60972 & 104& \mathrm{C}  \\     
185           &04\ 54\ 16.89 ,+02\ 54\ 24.8&       20.25&  62827 & 103& \mathrm{C}  \\     
\textit{186}  &04\ 54\ 16.94 ,+02\ 48\ 37.4&       20.03&  82173 &  96& \mathrm{T}  \\     
187           &04\ 54\ 17.10 ,+03\ 01\ 49.5&       19.35&  60006 &  57& \mathrm{T}  \\     
188           &04\ 54\ 17.31 ,+02\ 53\ 12.0&       19.07&  60264 &  60& \mathrm{T}  \\     
189           &04\ 54\ 17.33 ,+02\ 56\ 46.1&       19.60&  64056 & 122& \mathrm{P}  \\     
190           &04\ 54\ 17.43 ,+02\ 59\ 24.0&       19.54&  58780 &  76& \mathrm{C}  \\     
\textit{191}  &04\ 54\ 17.66 ,+02\ 48\ 24.9&       19.66&  82097 &  78& \mathrm{T}  \\     
192           &04\ 54\ 17.90 ,+02\ 55\ 35.0&       19.31&  60549 & 117& \mathrm{C}  \\     
193           &04\ 54\ 17.95 ,+02\ 46\ 49.6&       18.94&  60927 &  72& \mathrm{T}  \\     
194           &04\ 54\ 18.02 ,+02\ 57\ 41.5&       20.14&  61397 & 126& \mathrm{C}  \\     
195           &04\ 54\ 18.18 ,+02\ 59\ 55.7&       20.09&  61334 & 130& \mathrm{C}  \\     
196           &04\ 54\ 18.58 ,+03\ 00\ 36.5&       20.45&  57385 &  90& \mathrm{T}  \\     
\textit{197}  &04\ 54\ 18.88 ,+02\ 50\ 54.4&       16.93&  18754 & 100& \mathrm{T}  \\     
198           &04\ 54\ 19.00 ,+02\ 56\ 17.2&       18.74&  62504 & 135& \mathrm{C}  \\   
\textit{199}  &04\ 54\ 19.05 ,+02\ 56\ 13.8&       20.94&  91532 & 130& \mathrm{C}  \\   
200           &04\ 54\ 19.16 ,+02\ 58\ 26.5&       19.10&  60552 & 117& \mathrm{C}  \\     
201           &04\ 54\ 19.28 ,+03\ 01\ 09.9&       20.64&  58353 & 122& \mathrm{T+C}\\   
202           &04\ 54\ 19.31 ,+02\ 51\ 47.5&       19.71&  58948 &  60& \mathrm{T}  \\   
\textit{203}  &04\ 54\ 19.51 ,+02\ 48\ 05.7&       19.35&   8555 &  31& \mathrm{T}  \\     
204           &04\ 54\ 19.91 ,+02\ 57\ 44.8&       16.93&  60315 &  64& \mathrm{C+P}\\     
205           &04\ 54\ 19.96 ,+02\ 55\ 30.6&       17.35&  58597 &  99& \mathrm{C}  \\     
206           &04\ 54\ 20.17 ,+02\ 55\ 32.5&       19.43&  58381 & 108& \mathrm{C}  \\   
207           &04\ 54\ 20.21 ,+02\ 59\ 20.9&       20.88&  59955 & 176& \mathrm{C}  \\     
\textit{208}  &04\ 54\ 20.56 ,+03\ 00\ 55.9&       19.94&  76010 & 123& \mathrm{T+C}\\   
209           &04\ 54\ 20.58 ,+02\ 53\ 37.4&       19.52&  59946 &  90& \mathrm{T+C}\\     
210           &04\ 54\ 20.62 ,+02\ 56\ 41.4&       19.85&  58657 &  81& \mathrm{C}  \\   

              \noalign{\smallskip}			    
            \hline					    
            \noalign{\smallskip}			    
            \hline					    
         \end{array}\\
     $$ 
         \end{table}

\addtocounter{table}{-1}
\begin{table}[!ht]
          \caption[ ]{Continued.}
     $$ 
           \begin{array}{r c c r r c}

            \hline
            \noalign{\smallskip}
            \hline
            \noalign{\smallskip}

\mathrm{ID} & \mathrm{\alpha},\mathrm{\delta}\,(\mathrm{J}2000)  & r^\prime & \mathrm{v}\,\,\,\,\,\,\, & \mathrm{\Delta}\mathrm{v}& \mathrm{Source} \\
 & & &\mathrm{(\,km}&\mathrm{s^{-1}\,)}& \\
            \hline
            \noalign{\smallskip} 

211           &04\ 54\ 20.68 ,+02\ 55\ 29.8&       18.32&  58969 & 104& \mathrm{C}  \\     
212           &04\ 54\ 21.07 ,+02\ 51\ 24.9&       19.08&  61484 &  72& \mathrm{T}  \\     
213           &04\ 54\ 21.72 ,+02\ 55\ 56.0&       20.03&  58244 & 140& \mathrm{C}  \\     
214           &04\ 54\ 21.73 ,+03\ 05\ 11.6&       19.79&  61250 &  74& \mathrm{T}  \\     
215           &04\ 54\ 21.84 ,+02\ 55\ 00.0&       19.78&  58982 & 100& \mathrm{T}  \\     
\textit{216}  &04\ 54\ 22.02 ,+02\ 57\ 43.3&       22.27& 110602 & 225& \mathrm{C}  \\     
\textit{217}  &04\ 54\ 23.08 ,+02\ 50\ 18.9&       20.93& 165148 &  88& \mathrm{T}  \\     
218           &04\ 54\ 23.13 ,+02\ 58\ 01.2&       20.03&  60228 & 130& \mathrm{C}  \\     
\textit{219}  &04\ 54\ 23.27 ,+02\ 57\ 09.0&        -.- &  17085 & 108& \mathrm{C}  \\     
220           &04\ 54\ 23.27 ,+02\ 59\ 13.0&       19.83&  61202 &  94& \mathrm{C}  \\     
\textit{221}  &04\ 54\ 23.48 ,+03\ 03\ 16.1&       20.94&  78236 & 238& \mathrm{T}  \\     
222           &04\ 54\ 23.53 ,+02\ 50\ 34.5&       18.50&  60667 &  69& \mathrm{T}  \\     
\textit{223}  &04\ 54\ 23.84 ,+02\ 51\ 08.6&       20.14& 114162 &  98& \mathrm{T}  \\     
\textit{224}  &04\ 54\ 23.99 ,+02\ 56\ 10.9&       21.16&  70466 & 108& \mathrm{C}  \\     
\textit{225}  &04\ 54\ 24.24 ,+03\ 01\ 11.9&       21.50&  79751 & 117& \mathrm{T}  \\     
226           &04\ 54\ 24.88 ,+02\ 58\ 56.0&       19.93&  60289 &  87& \mathrm{C+P}\\     
227           &04\ 54\ 24.95 ,+02\ 52\ 23.5&       20.02&  60260 &  90& \mathrm{T}  \\     
228           &04\ 54\ 25.34 ,+02\ 47\ 03.9&       19.70&  61235 &  72& \mathrm{T}  \\     
229           &04\ 54\ 25.37 ,+02\ 49\ 23.4&       20.62&  57854 & 226& \mathrm{T}  \\     
230           &04\ 54\ 25.50 ,+02\ 59\ 38.3&       18.98&  60057 & 104& \mathrm{C}  \\     
\textit{231}  &04\ 54\ 26.18 ,+02\ 54\ 23.3&       21.25&  50350 & 135& \mathrm{C}  \\     
232           &04\ 54\ 26.63 ,+03\ 00\ 46.9&       19.87&  60182 &  77& \mathrm{T+C}\\     
\textit{233}  &04\ 54\ 26.80 ,+02\ 58\ 21.9&       19.86&  78677 & 135& \mathrm{C}  \\   
234           &04\ 54\ 27.64 ,+03\ 03\ 29.3&       20.02&  60186 &  99& \mathrm{T}  \\   
\textit{235}  &04\ 54\ 27.76 ,+02\ 55\ 29.2&       18.63&  67171 &  90& \mathrm{C}  \\     
236           &04\ 54\ 27.96 ,+02\ 54\ 18.1&       19.61&  61070 &  93& \mathrm{T}  \\      
237           &04\ 54\ 28.14 ,+02\ 55\ 45.7&       20.41&  59611 & 112& \mathrm{C}  \\   
\textit{238}  &04\ 54\ 28.18 ,+02\ 55\ 36.5&       18.74&  66893 &  99& \mathrm{C}  \\     
239           &04\ 54\ 28.63 ,+03\ 04\ 16.0&       19.17&  60508 &  62& \mathrm{T}  \\     
\textit{240}  &04\ 54\ 29.02 ,+02\ 54\ 29.0&       19.97&  65186 &  61& \mathrm{T+C}\\     
\textit{241}  &04\ 54\ 29.02 ,+02\ 56\ 59.5&       20.54&  62384 & 153& \mathrm{C}  \\   
\textit{242}  &04\ 54\ 29.14 ,+02\ 48\ 55.9&       19.74& 135190 & 366& \mathrm{T}  \\     
\textit{243}  &04\ 54\ 29.53 ,+02\ 58\ 22.1&       19.34&  67096 & 108& \mathrm{C}  \\   
244           &04\ 54\ 29.58 ,+02\ 55\ 22.0&       19.66&  62063 & 112& \mathrm{C}  \\     
\textit{245}  &04\ 54\ 30.21 ,+03\ 02\ 25.3&       18.59&  66861 &  98& \mathrm{T}  \\   

              \noalign{\smallskip}			    
            \hline					    
            \noalign{\smallskip}			    
            \hline					    
         \end{array}\\
     $$ 
         \end{table}

\addtocounter{table}{-1}
\begin{table}[!ht]
          \caption[ ]{Continued.}
     $$ 
           \begin{array}{r c c r r c}

            \hline
            \noalign{\smallskip}
            \hline
            \noalign{\smallskip}

\mathrm{ID} & \mathrm{\alpha},\mathrm{\delta}\,(\mathrm{J}2000)  & r^\prime & \mathrm{v}\,\,\,\,\,\,\, & \mathrm{\Delta}\mathrm{v}& \mathrm{Source} \\
 & & &\mathrm{(\,km}&\mathrm{s^{-1}\,)}& \\
            \hline
            \noalign{\smallskip} 

246           &04\ 54\ 30.31 ,+02\ 58\ 44.8&       19.18&  59877 & 104& \mathrm{C}  \\     
\textit{247}  &04\ 54\ 30.67 ,+02\ 54\ 44.4&       20.05&  66872 & 117& \mathrm{C}  \\     
\textit{248}  &04\ 54\ 30.90 ,+02\ 59\ 35.4&       19.93&  17451 & 225& \mathrm{C}  \\     
249           &04\ 54\ 31.01 ,+02\ 49\ 04.3&       20.05&  60211 &  64& \mathrm{T}  \\     
250           &04\ 54\ 31.23 ,+03\ 05\ 13.6&       18.97&  60001 &  88& \mathrm{T}  \\     
\textit{251}  &04\ 54\ 31.42 ,+02\ 57\ 22.9&       20.36&  74618 & 180& \mathrm{C}  \\     
\textit{252}  &04\ 54\ 31.93 ,+02\ 52\ 36.7&       20.39&  76231 &  41& \mathrm{T+C}\\     
253           &04\ 54\ 32.31 ,+03\ 03\ 52.1&       20.23&  59672 & 147& \mathrm{T}  \\     
254           &04\ 54\ 32.63 ,+02\ 53\ 01.3&       20.32&  59221 & 164& \mathrm{T}  \\     
255           &04\ 54\ 32.68 ,+02\ 54\ 48.9&       18.00&  60894 &  59& \mathrm{C+P}\\     
\textit{256}  &04\ 54\ 33.56 ,+03\ 03\ 23.1&       19.10&  99246 & 112& \mathrm{C}  \\     
\textit{257}  &04\ 54\ 33.78 ,+02\ 58\ 51.8&       20.91& 125307 & 225& \mathrm{C}  \\     
\textit{258}  &04\ 54\ 34.25 ,+02\ 50\ 00.3&       19.46& 111247 &  81& \mathrm{T}  \\     
259           &04\ 54\ 35.27 ,+03\ 01\ 05.3&       18.84&  59269 & 104& \mathrm{C}  \\     
260           &04\ 54\ 35.83 ,+03\ 01\ 05.4&       20.53&  61005 & 153& \mathrm{C}  \\     
\textit{261}  &04\ 54\ 36.49 ,+02\ 54\ 33.7&       21.45& 112128 & 225& \mathrm{C}  \\     
262           &04\ 54\ 36.93 ,+03\ 03\ 23.7&       19.63&  60798 & 122& \mathrm{C}  \\     
263           &04\ 54\ 37.47 ,+03\ 02\ 24.4&       18.65&  60021 & 112& \mathrm{C}  \\     
264           &04\ 54\ 38.54 ,+03\ 00\ 51.5&       17.43&  61074 & 104& \mathrm{C}  \\     
265           &04\ 54\ 39.39 ,+03\ 03\ 44.7&       20.44&  60825 & 189& \mathrm{C}  \\     
\textit{266}  &04\ 54\ 40.49 ,+02\ 52\ 31.8&       19.67&  66941 & 144& \mathrm{C}  \\     
267           &04\ 54\ 40.76 ,+03\ 02\ 35.9&       19.51&  60006 & 117& \mathrm{C}  \\     
\textit{268}  &04\ 54\ 41.11 ,+02\ 59\ 49.5&       21.97& 113621 & 450& \mathrm{C}  \\   
\textit{269}  &04\ 54\ 41.67 ,+02\ 59\ 17.8&       20.89& 106558 & 108& \mathrm{C}  \\   
\textit{270}  &04\ 54\ 41.90 ,+02\ 58\ 55.2&       19.60&  69483 & 270& \mathrm{C}  \\     
\textit{271}  &04\ 54\ 42.25 ,+03\ 02\ 02.1&       20.33&  99066 & 126& \mathrm{C}  \\   
272           &04\ 54\ 42.63 ,+03\ 01\ 58.3&       19.24&  59602 & 144& \mathrm{C}  \\   
\textit{273}  &04\ 54\ 43.01 ,+02\ 57\ 42.3&       20.60&  70781 &  72& \mathrm{C}  \\     
\textit{274}  &04\ 54\ 44.96 ,+03\ 01\ 06.2&       20.78& 113447 & 225& \mathrm{C}  \\     
\textit{275}  &04\ 54\ 45.34 ,+02\ 53\ 40.4&       19.59&  55623 & 104& \mathrm{C}  \\     
\textit{276}  &04\ 54\ 46.28 ,+02\ 52\ 36.1&       20.56&  77385 &  99& \mathrm{C}  \\   
\textit{277}  &04\ 54\ 46.85 ,+02\ 53\ 21.9&       20.84&  55396 &  99& \mathrm{C}  \\     
\textit{278}  &04\ 54\ 46.85 ,+03\ 02\ 57.0&       21.49&  98017 &  99& \mathrm{C}  \\   
\textit{279}  &04\ 54\ 47.88 ,+03\ 03\ 33.3&       18.62&  71557 & 112& \mathrm{C}  \\     
280           &04\ 54\ 48.86 ,+02\ 52\ 34.5&       20.66&  62366 & 180& \mathrm{C}  \\   

              \noalign{\smallskip}			    
            \hline					    
            \noalign{\smallskip}			    
            \hline					    
         \end{array}\\
     $$ 
         \end{table}

\addtocounter{table}{-1}
\begin{table}[!ht]
          \caption[ ]{Continued.}
     $$ 
           \begin{array}{r c c r r c}

            \hline
            \noalign{\smallskip}
            \hline
            \noalign{\smallskip}

\mathrm{ID} & \mathrm{\alpha},\mathrm{\delta}\,(\mathrm{J}2000)  & r^\prime & \mathrm{v}\,\,\,\,\,\,\, & \mathrm{\Delta}\mathrm{v}& \mathrm{Source} \\
 & & &\mathrm{(\,km}&\mathrm{s^{-1}\,)}& \\
            \hline
            \noalign{\smallskip} 

\textit{281}  &04\ 54\ 49.21 ,+03\ 03\ 23.4&       20.20&  99588 & 180& \mathrm{C}  \\     
\textit{282}  &04\ 54\ 51.40 ,+02\ 57\ 52.4&       19.39&  45940 &  99& \mathrm{C}  \\     
\textit{283}  &04\ 54\ 52.12 ,+02\ 54\ 27.4&       20.81&  69405 & 117& \mathrm{C}  \\     
\textit{284}  &04\ 54\ 52.32 ,+02\ 56\ 42.7&       20.45& 110251 & 225& \mathrm{C}  \\     
285           &04\ 54\ 53.87 ,+02\ 52\ 45.2&       20.03&  60537 & 126& \mathrm{C}  \\     
286           &04\ 54\ 55.00 ,+02\ 54\ 00.0&       19.21&  62489 & 148& \mathrm{C}  \\     
\textit{287}  &04\ 54\ 55.47 ,+02\ 58\ 31.7&       22.45& 138135 & 144& \mathrm{C}  \\     
\textit{288}  &04\ 54\ 56.66 ,+03\ 00\ 53.4&       21.63&  94455 & 112& \mathrm{C}  \\     
\textit{289}  &04\ 54\ 56.87 ,+02\ 54\ 02.9&       18.69&  69321 &  99& \mathrm{C}  \\     
\textit{290}  &04\ 54\ 57.20 ,+03\ 03\ 26.6&       20.64&  98625 & 104& \mathrm{C}  \\     
\textit{291}  &04\ 54\ 58.37 ,+03\ 01\ 54.0&       20.73& 135512 & 225& \mathrm{C}  \\     
\textit{292}  &04\ 54\ 59.89 ,+02\ 53\ 31.1&       19.46&  68559 & 135& \mathrm{C}  \\     
\textit{293}  &04\ 55\ 01.66 ,+02\ 57\ 55.7&        -.- & 138036 & 225& \mathrm{C}  \\     

              \noalign{\smallskip}			    
            \hline					    
            \noalign{\smallskip}			    
            \hline					    
         \end{array}\\
     $$ 
         \end{table}


In summary, our redshift catalog of A520 consists of 293 galaxies
sampling a wide, asymmetric cluster region (see Fig.~\ref{figottico})
and having a median error on $cz$ of 112 \kss.
Table~\ref{catalogA520} lists the velocity catalog: identification
number of each galaxy, ID (Col.~1); right ascension and declination,
$\alpha$ and $\delta$ (J2000, Col.~2); $r'$ magnitudes (Col.~3);
heliocentric radial velocities, ${\rm v}=cz_{\sun}$ (Col.~4) with
errors, $\Delta {\rm v}$ (Col.~5); redshift source (Col.~6; T:TNG,
C:CNOC and P:Proust et al.).  We list $r'$ magnitudes for 291 out of
293 galaxies having redshifts. The exceptions are a galaxy just
outside the western border of the imaging field and a huge foreground
spiral galaxy.  We have redshifts for galaxies down to
$r^\prime\sim$21.5 mag, but we are 40\% complete down to $r^\prime$=19
mag within 3 arcmin from
R.A.=$04^{\mathrm{h}}54^{\mathrm{m}}14^{\mathrm{s}}$, Dec.=$+02\degree
57\arcmm 00\arcsec$ (J2000.0). The completeness of the spectroscopic
sample decreases in the outskirts of the cluster.

\begin{figure}
\centering 
\resizebox{\hsize}{!}{\includegraphics{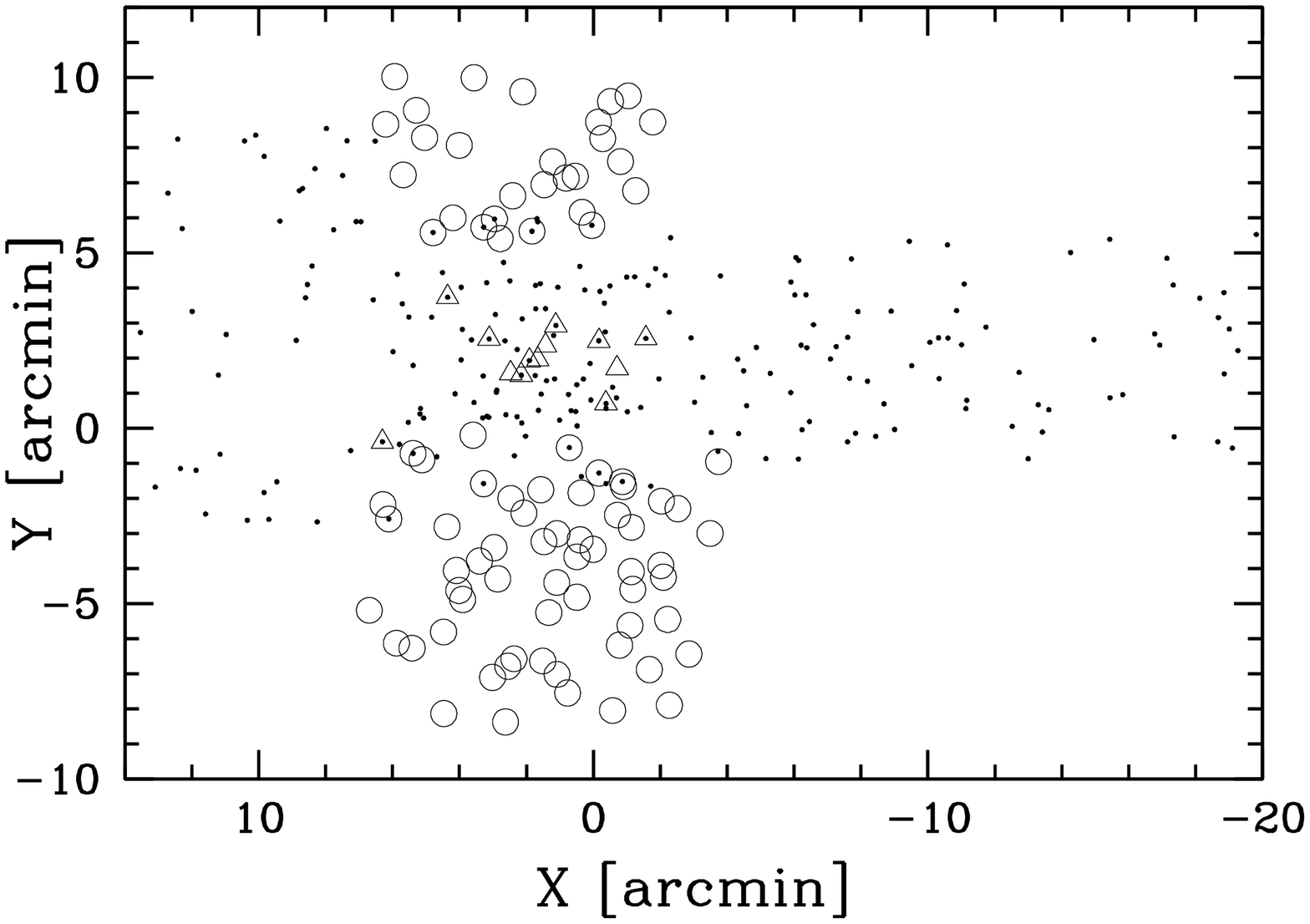}}
\vspace{-3cm}
\caption
{Spatial distribution on the sky of the 293 galaxies having redshifts
  in the cluster field.  
Circles indicate galaxies
  having new redshifts acquired with the TNG. Dots and triangles
  indicate galaxies having redshift data from CNOC and Proust et
  al. (\cite{pro00}) catalogs, respectively. 
The X--ray peak is taken as the cluster
  center.  
}
\label{figcat}
\end{figure}

Figure~\ref{figcat} shows the contribute of TNG data
added to previous spectroscopic information.

A520 does not exhibit the presence of a clear dominant galaxy and in
fact it is classified as Bautz--Morgan class III (Abell et
al. \cite{abe89}). In particular, our sample lists nine luminous
galaxies in a range of one mag from the most luminous one: IDs~204, 106,
44, 170, 95, 205, 264, 160 and 137. These galaxies are generally
sparse in the field. A few of these galaxies are close to the lensing
mass peaks pointed out by M07, i.e. ID~204 is close to peak No. 1; the
galaxy couple composed by IDs~160 and 170 is close to peak No. 2;
ID~106 is close to peak No. 4; ID~205 is close to peak
No. 5. 

Govoni et al. (\cite{gov01b}) pointed out the presence of several
discrete radio sources in the field of A520. In particular, there are
two head--tail radio sources (0454+0255A and 0454+0255B; see also
Cooray et al. \cite{coo98}) located on the eastern side with the tails
oriented toward the same direction, opposite to the cluster center.
Our catalog lists the redshift for the northern one, 0454+0255B
(ID~184), which is classified as a cluster member. Cooray et
al. (\cite{coo98}) also list a third radio source (0454+0257) which,
again, is classified as a cluster member (ID~95). From a visual
inspection of the Chandra image studied by Markevitch et
al. (\cite{mar05}, Obs.Id 4215) we also note that 0454+0257 is an
evident pointlike X--ray source in the field of A520.

\section{Analysis and results}
\label{anal}

\subsection{Member selection}
\label{memb}

\begin{figure}
\centering
\resizebox{\hsize}{!}{\includegraphics{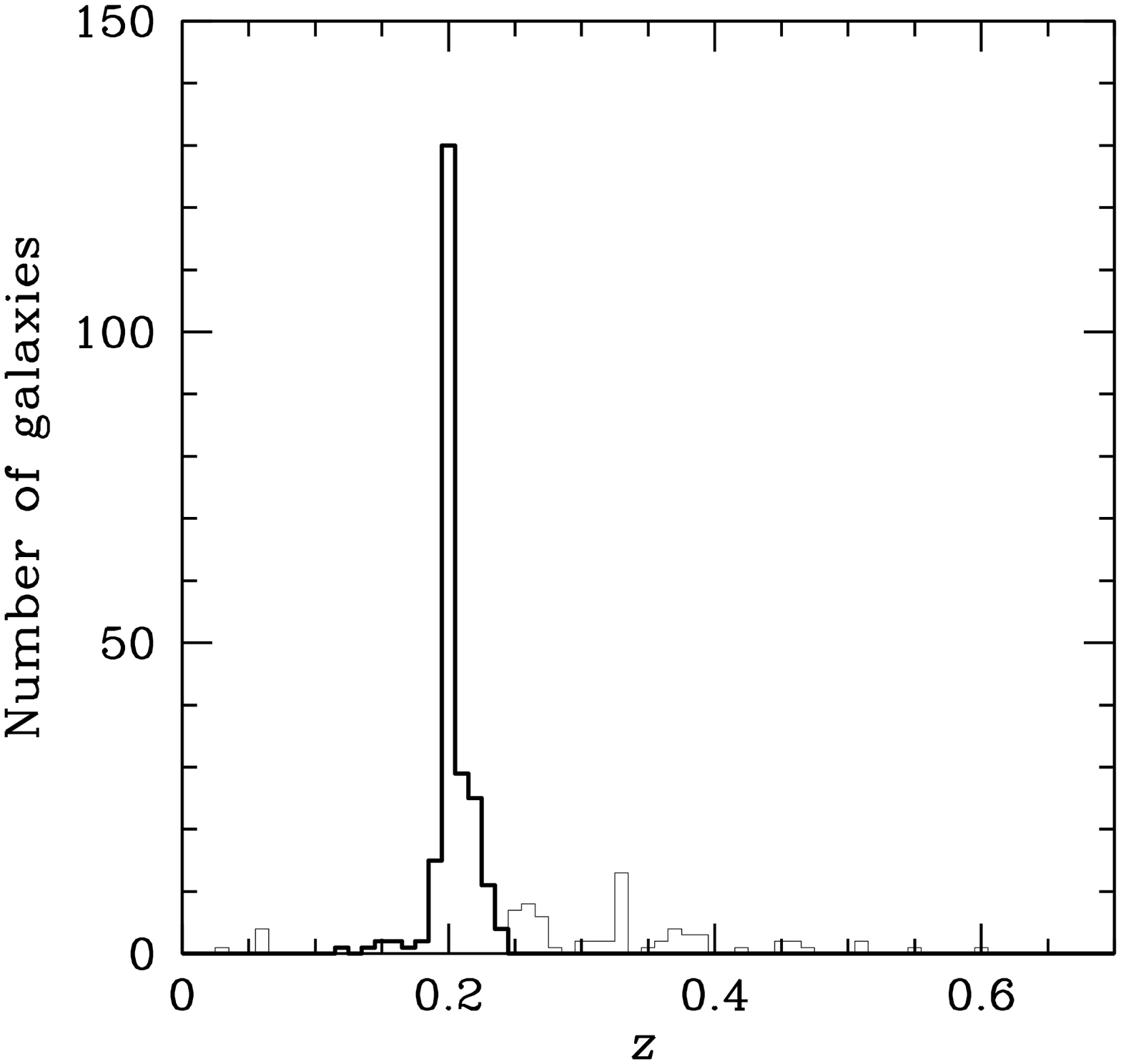}}
\caption
{Redshift galaxy distribution. The solid line histogram refers to
 the (223) galaxies assigned to the cluster according to the DEDICA
  reconstruction method.}
\label{fighisto}
\end{figure}

To select cluster members out of 293 galaxies having redshifts, we
follow a two steps procedure.  First, we perform the adaptive--kernel
method (hereafter DEDICA, Pisani \cite{pis93} and \cite{pis96}; see
also Fadda et al. \cite{fad96}; Girardi et al. \cite{gir96}; Girardi
\& Mezzetti \cite{gir01}). We search for significant peaks in the
velocity distribution at $>$99\% c.l.. This procedure detects A520 as
an asymmetric one--peak structure at $z\sim0.201$ populated by 223
galaxies considered as candidate cluster members (see
Fig.~\ref{fighisto}).

\begin{figure}
\centering 
\resizebox{\hsize}{!}{\includegraphics{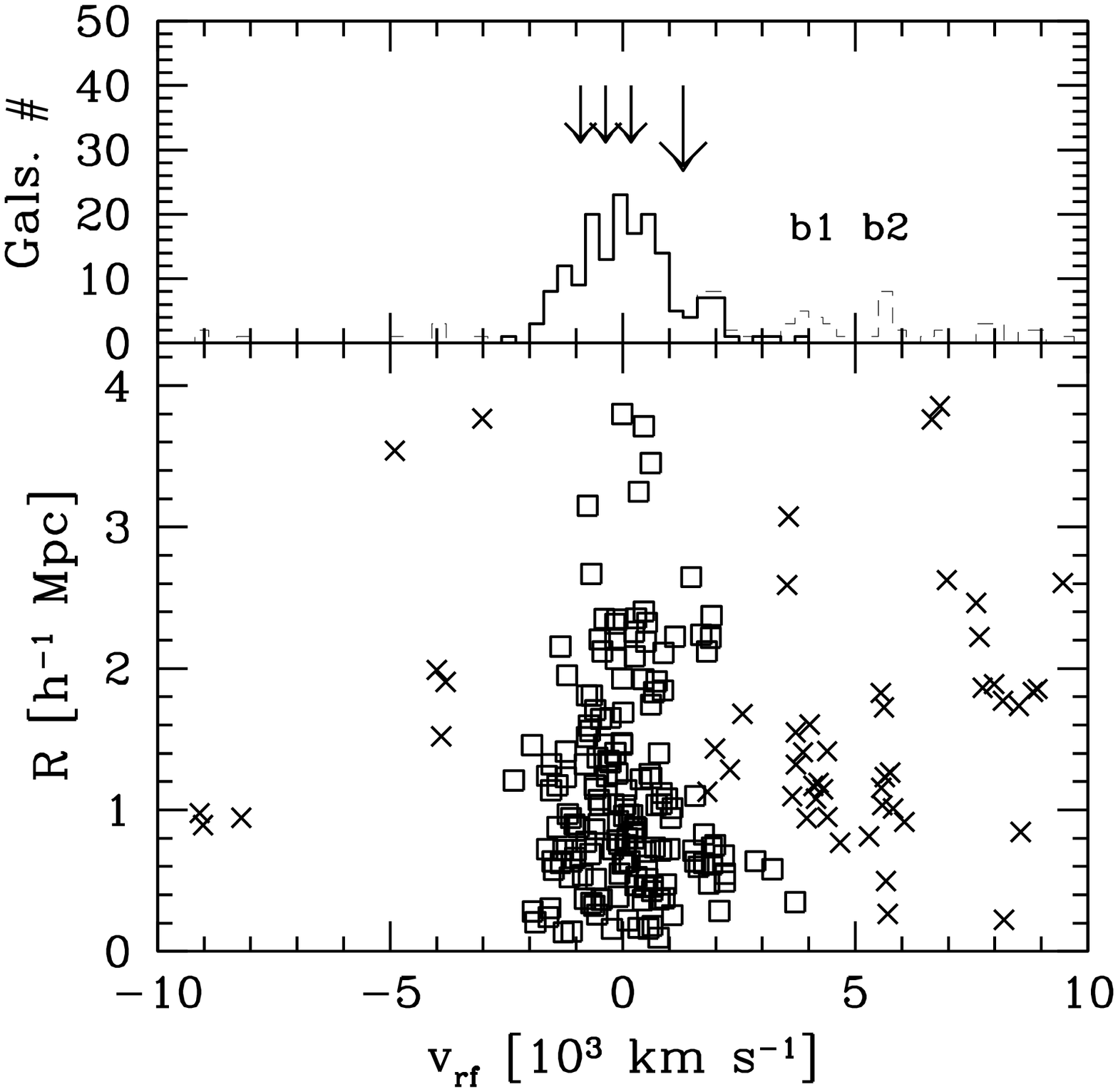}}
\caption
{{\em Lower panel}: projected clustercentric distance vs. rest--frame
  velocity for the 223 galaxies in the main peak
  (Fig.~\ref{fighisto}).  Crosses show galaxies detected as
  interlopers by our ``shifting gapper'' procedure.  {\em Upper
    panel}: rest--frame velocity histogram for the 223 galaxies in the
  main peak; the solid line refers to the 167 cluster members
  only. Large and small arrows indicate the positions of weighted gaps
  in the velocity distribution of the whole sample and of the main
  system (MS).  Labels b1 and b2 indicate back1 and back2 ``background''
  peaks of galaxies.}
\label{figvd}
\end{figure}

All the galaxies assigned to the A520 peak are analyzed in the second
step which uses the combination of position and velocity information:
the ``shifting gapper'' method by Fadda et al. (\cite{fad96}).  This
procedure rejects galaxies that are too far in velocity from the main
body of galaxies and within a fixed bin that shifts along the distance
from the cluster center.  The procedure is iterated until the number
of cluster members converges to a stable value.  Following Fadda et
al. (\cite{fad96}) we use a gap of $1000$ \ks -- in the cluster
rest--frame -- and a bin of 0.6 \hh, or large enough to include 15
galaxies.  The choice of the cluster center is not obvious.  In fact,
several galaxy condensations are visible in the field (Gioia \&
Luppino \cite{gio94}). Moreover, no obvious dominant galaxy is present
(see Sect.~\ref{galc}) and the lensing mass distribution shows several
peaks (e.g., M07). Thus, hereafter we assume the position of the peak
of X--ray emission as listed by Ebeling et al. (\cite{ebe96})
[R.A.=$04^{\mathrm{h}}54^{\mathrm{m}}07\dotsec44$, Dec.=$+02\degree
  55\arcmm 12\dotarcs0$ (J2000.0)] as the cluster center. After the
``shifting gapper'' procedure we obtain a sample of 167 fiducial
cluster members (see Fig.~\ref{figvd}).

The 2D galaxy distribution analyzed through the 2D DEDICA method shows
only one peak [at R.A.=$04^{\mathrm{h}}54^{\mathrm{m}}13\dotsec55$,
  Dec.=$+02\degree 56\arcmm 35\dotarcs2$ (J2000.0)]. This peak,
hereafter the ``optical'' cluster center, is displaced towards NE with
respect to the X--ray peak and is close, but not coincident, to a pair
of luminous galaxies (IDs 160 and 170). The biweight cluster center,
i.e. that recovered by computing the biweight means (Beers et
al. \cite{bee90}) of R.A. and Dec. of galaxy positions
[R.A.=$04^{\mathrm{h}}54^{\mathrm{m}}12\dotsec62$, Dec.=$+02\degree
  55\arcmm 57\dotarcs3$ (J2000.0)], is roughly coincident with the
DEDICA peak. Using these alternative cluster centers we verify the
robustness of our member selection.

\subsection{Global kinematical properties}
\label{glob}

By applying the biweight estimator to the 167 cluster members (Beers
et al. \cite{bee90}), we compute a mean cluster redshift of
$\left<z\right>=0.2008\pm$ 0.0003, i.e.
$\left<\rm{v}\right>=(60209\pm$82) \kss.  We estimate the LOS velocity
dispersion, $\sigma_{\rm v}$, by using the biweight estimator and
applying the cosmological correction and the standard correction for
velocity errors (Danese et al. \cite{dan80}).  We obtain $\sigma_{\rm
  v}=1066_{-61}^{+67}$ \kss, where errors are estimated through a
bootstrap technique.

\begin{table}
        \caption[]{Results of the weighted gap analysis for the whole sample and for the MS subsystem.}
         \label{tabgap}
                $$
         \begin{array}{l c c c c}
            \hline
            \noalign{\smallskip}
            \hline
            \noalign{\smallskip}
\mathrm{Sample} & \mathrm{N_{gals\ pre},N_{gals\ aft}} & 
\mathrm{v_{pre},v_{aft}}& 
\mathrm{Size}& \mathrm{Prob.}\\
                &                  &\mathrm{km\ s^{-1}}&              &
         \\
            \hline
            \noalign{\smallskip}
 
\mathrm{Whole\ Sample} & 145,22 &61547,61964&3.81& 5.0E-4\\
\mathrm{MS}  & 30,30  &59059,59163&2.52& 1.4E-2\\
\mathrm{MS}  & 30,33  &59722,59783&2.35& 3.0E-2\\
\mathrm{MS}  & 33,52  &60440,61548&2.34& 3.0E-2\\
              \noalign{\smallskip}
            \hline
            \noalign{\smallskip}
            \hline
         \end{array}
$$
         \end{table}

\begin{figure}
\centering
\resizebox{\hsize}{!}{\includegraphics{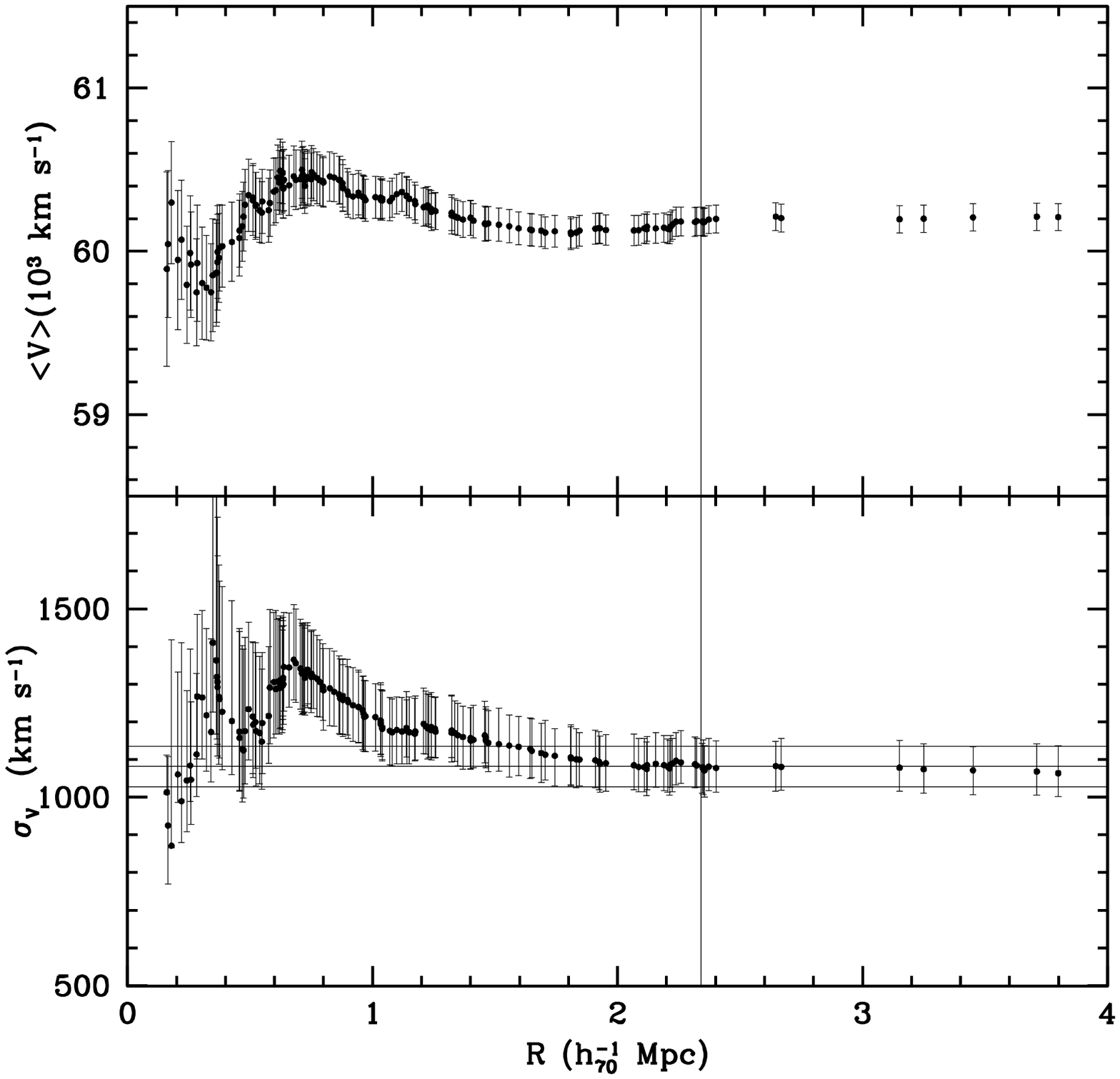}}
\caption
{Integral profiles of mean velocity ({\em upper panel}) and LOS
  velocity dispersion ({\em lower panel}). The mean and dispersion at
  a given (projected) radius from the cluster center is estimated by
  considering all galaxies within that radius (the first point is
  obtained on the basis of the five galaxies close to the cluster
  center). The error bands at the $68\%$ c.l. are shown.  In the lower
  panel, the horizontal line represents the X--ray temperature with
  the respective 90 per cent errors (Govoni et al. \cite{gov04})
  transformed in $\sigma_{\rm v}$ assuming the density--energy
  equipartition between gas and galaxies, i.e.  $\beta_{\rm spec}=1$
  (see text).}
\label{figprof}
\end{figure}

To evaluate the robustness of the $\sigma_{\rm v}$ estimate we analyze
the velocity dispersion profile (Fig.~\ref{figprof}).  The integral
profile smoothly decreases and flattens beyond $\sim0.6$ \h suggesting
that a robust value of $\sigma_{\rm v}$ is asymptotically reached in
the external cluster regions, as found for most nearby clusters (e.g.,
Fadda et al. \cite{fad96}; Girardi et al. \cite{gir96}).

\subsection{Substructure}
\label{clus}

\subsubsection{Velocity distribution}
\label{velo}

We analyze the velocity distribution to look for possible deviations
from Gaussianity that might provide important signatures of complex
dynamics. For the following tests the null hypothesis is that the
velocity distribution is a single Gaussian.

We estimate three shape estimators, i.e. the kurtosis, the skewness,
and the scaled tail index (see, e.g., Beers et al.~\cite{bee91}).
According to the value of the skewness (+0.471) the velocity
distribution is positively skewed and differs from a Gaussian at the
$95-99\%$ c.l. (see Table~2 of Bird \& Beers \cite{bir93}). Moreover,
according the the scaled tail index the velocity distribution is
heavily tailed and differs from a Gaussian at the $90-95\%$ c.l. (see
Table~2 of Bird \& Beers~\cite{bir93}).

Then we investigate the presence of gaps in the velocity distribution.
A weighted gap in the space of the ordered velocities is defined as
the difference between two contiguous velocities, weighted by the
location of these velocities with respect to the middle of the
data. We obtain values for these gaps relative to their average size,
precisely the midmean of the weighted--gap distribution. We look for
normalized gaps larger than 2.25 since in random draws of a Gaussian
distribution they arise at most in about $3\%$ of the cases,
independent of the sample size (Wainer and Schacht~\cite{wai78}; see
also Beers et al.~\cite{bee91}). We detect a significant gap (at the
$99.95\%$ c.l.)  which separates the main cluster from a group of 22
high velocity galaxies (see Fig.~\ref{figvd} and the first line of
Table~\ref{tabgap}). For each gap Table~\ref{tabgap} lists the number
of galaxies for the group before the gap and that after the gap
(Col.~2); the velocity boundaries before and after the gap (Col.~3);
the size of the gap (Col.~4); the probability of finding such a gap in
a Gaussian distribution (Col.~5). Hereafter we define MS the main
system with the 145 galaxies having low velocities and HGV the group
with the 22 galaxies having high velocities (see Table~\ref{tabv} for
their main kinematical properties).

\begin{table}
        \caption[]{Global properties of the whole sample, the MS and
the HVG.}
         \label{tabv}
                $$
         \begin{array}{l r l l c c}
            \hline
            \noalign{\smallskip}
            \hline
            \noalign{\smallskip}
\mathrm{Sample} & \mathrm{N_g} & \phantom{249}\mathrm{<v>}\phantom{249} & 
\phantom{24}\sigma_{\rm v}\phantom{24}&\mathrm{R_{vir}}&\mathrm{Mass(<R_{vir})}\\
& &\phantom{249}\mathrm{km\ s^{-1}}\phantom{249}&\phantom{2}\mathrm{km\ s^{-1}}\phantom{24}&\mathrm{h_{70}^{-1}Mpc}&\mathrm{h_{70}^{-1}}10^{14}\mathrm{M}_{\odot}\\
            \hline
            \noalign{\smallskip}
\mathrm{Whole\ system}     &167 &60209\pm82&1066_{-61}^{+67} &2.34&17\pm2\\
\mathrm{MS}   & 145&59978\pm67& \phantom{1}812_{-46}^{+35} &1.79&\phantom{1}8\pm2\\
\mathrm{HVG}        & 22 &62419\pm74& \phantom{1}338_{-84}^{+225}&0.74&0.6_{-0.3}^{+0.8}\\
              \noalign{\smallskip}
            \hline
            \noalign{\smallskip}
            \hline
         \end{array}
$$
         \end{table}

\begin{figure}
\centering 
\resizebox{\hsize}{!}{\includegraphics{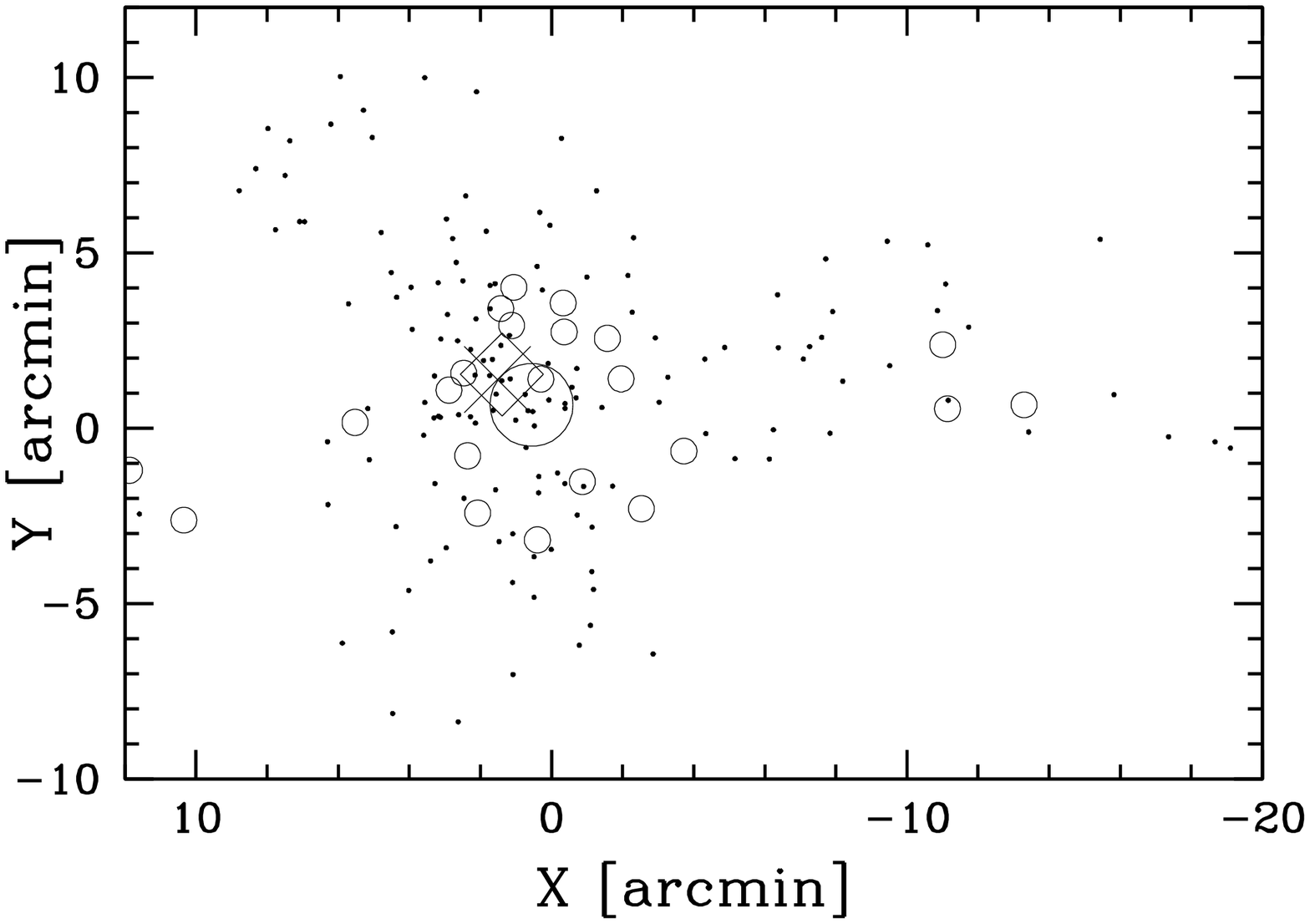}}
\vspace{-3cm}
\caption
{Spatial distribution on the sky of the 167 galaxies of the whole
  cluster showing the two groups recovered by the weighted gap
  analysis. Dots and small circles indicate the main system (MS) and
  the high velocity group (HVG) galaxies.  Large cross, rotated square
  and circle indicate the optical centers of the whole cluster, the MS
  and the HVG, respectively.  The X--ray peak is taken as the cluster
  center.  }
\label{figkmm}
\end{figure}

As for the spatial distribution, there is no difference between the
galaxies of the HVG and the MS (according to the 2D
Kolmogorov--Smirnov test -- hereafter 2DKS--test - Fasano \&
Franceschini \cite{fas87}, see Fig.~\ref{figkmm}).  However, when
considering the clustercentric distances, they differ at the $97.5\%$
c.l. according to the 1D Kolmogorov--Smirnov test (hereafter
1DKS--test, see e.g., Press et
al. \cite{pre92}) with the HVG galaxies being, on average, closer to
the cluster (X--ray) center.  Accordingly, while the optical
(biweight) center of the MS lies very close to the optical center of
the whole cluster, the optical (biweight) center of the HVG is closer
to the X--ray cluster center (see Fig.~\ref{figkmm}).

We also use the results of the gap analysis to determine the first
guess when using the Kaye's mixture model (KMM) test to find a
possible group partition of the velocity distribution (as implemented
by Ashman et al. \cite{ash94}). The KMM algorithm fits a
user--specified number of Gaussian distributions to a dataset and
assesses the improvement of that fit over a single Gaussian. In
addition, it provides the maximum--likelihood estimate of the unknown
n--mode Gaussians and an assignment of objects into groups.  We find a
two--groups partition which is a significantly better descriptor of the
velocity distribution with respect to a single Gaussian at the $95\%$
cl.. The cluster partition is similar to that indicated by the above
weighted gap analysis detecting two groups with 146 and 21 galaxies.

\subsubsection{Dressler--Shectman statistics}
\label{comb}

\begin{figure}
\centering 
\resizebox{\hsize}{!}{\includegraphics{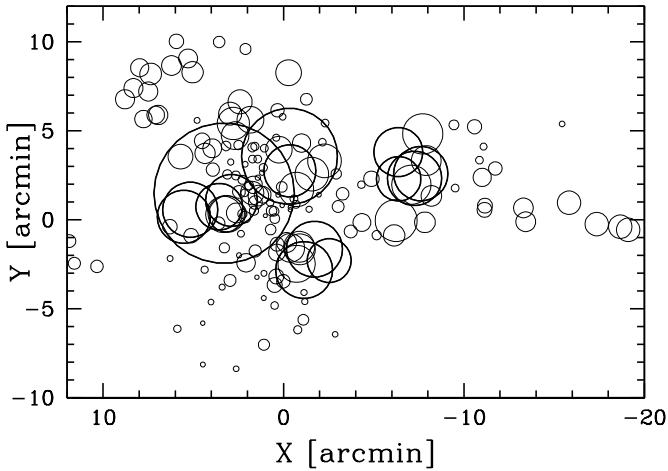}}
\vspace{-3cm}
\caption
{Spatial distribution of the 167 cluster members,
  each marked by a circle: the larger the circle, the larger is the
  deviation $\delta_i$ of the local parameters from the global cluster
  parameters, i.e. there is more evidence for substructure (according
  to the Dressler \& Shectman test, see text).  Heavy circles
  indicate those with $\delta_i \ge 2.5$.  
  }
\label{figds}
\end{figure}

We also analyze substructure combining galaxy velocity and position
information. We compute the $\Delta$--statistics devised by Dressler
\& Shectman (\cite{dre88}, hereafter DS).  We find a significant
indication of DS substructure (at the $97\%$ c.l. using 1000 Monte
Carlo simulations; see e.g. Boschin et al. \cite{bos04}).
Figure~\ref{figds} shows the distribution on the sky
of all galaxies, each marked by a circle: the larger the circle, the
larger the deviation $\delta_i$ of the local kinematical parameters
from the global cluster parameters, i.e. the higher the evidence for
substructure.

\begin{figure}
\centering 
\resizebox{\hsize}{!}{\includegraphics{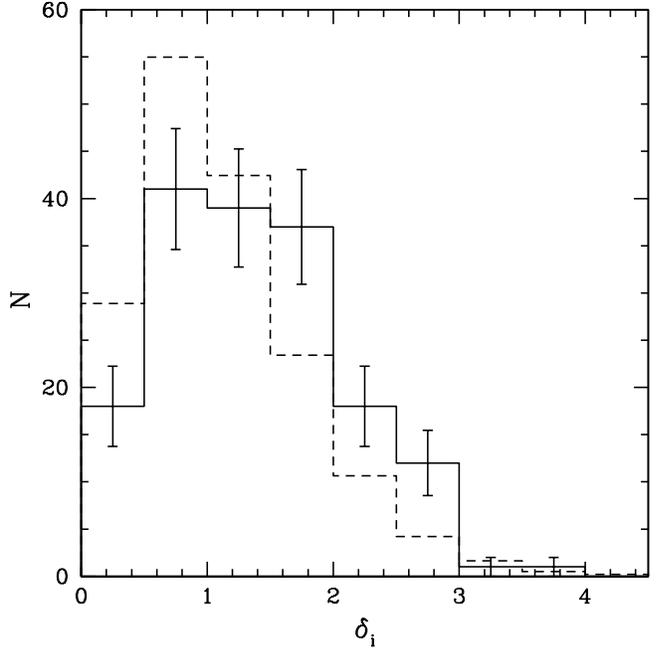}}
\caption
{The distribution of $\delta_i$ deviations of the Dressler--Shectman
  analysis for the 167 member galaxies. The solid line represents the
  observations, the dashed line the distribution for the galaxies of
  simulated clusters, normalized to the observed number.}
\label{figdeltai}
\end{figure}

To better point out galaxies belonging to substructures, we resort to
the technique developed by Biviano et al. (\cite{biv02}, see also
Boschin et al. \cite{bos06}; Girardi et al. \cite{gir06}), who used
the individual $\delta_i$--values of the DS method. The critical point
is to determine the value of $\delta_i$ that optimally indicates
galaxies belonging to substructure. To this aim we consider the
$\delta_i$--values of all 1000 Monte Carlo simulations used above.
The resulting distribution of $\delta_i$ is compared to the observed
one finding a difference at the $99\%$ c.l. according to the
1DKS--test.  The ``simulated'' distribution is normalized to produce
the observed number of galaxies and compared to the observed
distribution in Fig.~\ref{figdeltai}: the latter shows a tail at large
values.  Selecting galaxies with $\delta_i \le 2.5$ the 1DKS--test
gives only a marginal difference between real and simulated galaxies
(at the $93\%$ c.l.)  suggesting that galaxies with $\delta_i > 2.5$
presumably are in substructures. These galaxies are indicated with
heavy circles in Fig.~\ref{figds} showing four subclumps (at northern,
eastern, southern and distant western cluster regions).

\subsubsection{Analysis of velocity dispersion profiles}
\label{2d}

\begin{figure}
\centering 
\resizebox{\hsize}{!}{\includegraphics{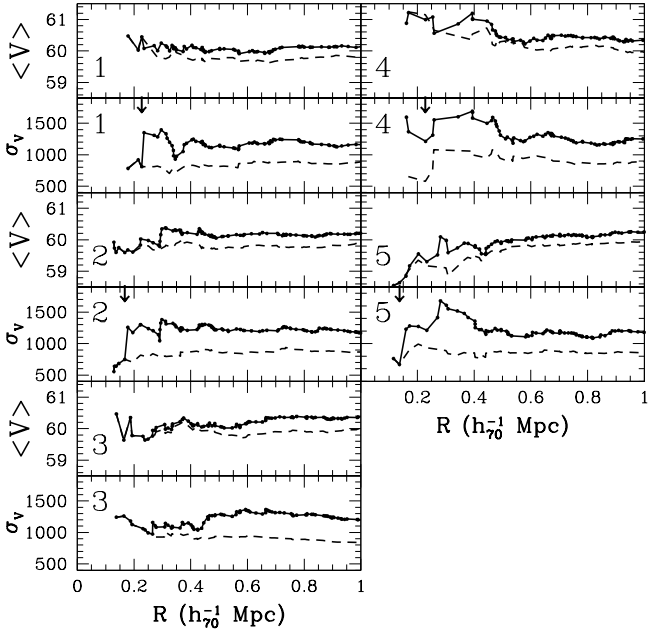}}
\caption
{Kinematical profiles of the galaxy clump surrounding the lensing mass
  peaks No. 1-5 listed by M07.  For each peak integral mean velocity
  (in units of $10^3$ \kss) and LOS velocity dispersion (in units of
  \kss) profiles are shown in upper and lower panels,
  respectively. The technique is the same adopted in
  Fig.~\ref{figprof}, but we omit errors for the sake of clarity.  The
  solid and dashed lines join the values obtained using all cluster
  galaxies and only galaxies belonging to the MS, respectively.  The arrows
  indicate the regions likely not contaminated from other subclumps
  (see Sect.~\ref{2d} and Sect.~\ref{2dms}).}
\label{figprofmul}
\end{figure}

Finally we analyze the kinematical properties of galaxy populations
located in different spatial regions of the cluster.  We compute the
profiles of mean velocity and velocity dispersion of galaxy systems
surrounding the lensing mass peaks listed by M07 (see
Fig.~\ref{figprofmul}).  This allows an independent analysis of the
possible individual galaxy clumps.  A quasi flat profile is expected
in the case of a relaxed system with isotropic orbits for galaxies
(e.g., Girardi et al. \cite{gir98}). Although an increasing/decreasing
profile might be due to particular orbits of galaxies in a relaxed
system (e.g., Girardi et al. \cite{gir98}; Biviano \& Katgert
\cite{biv04}), here this is likely connected with the presence of
substructure.  As for an increasing profile, this might be simply
induced by the contamination of the galaxies of a close, secondary
clump having a different mean velocity (e.g., Girardi et
al. \cite{gir96}; Girardi et al. \cite{gir05}).  This hypothesis can
be investigated by looking at the behavior of the mean velocity
profile.  In fact, if the $\sigma_{\rm v}$ profile increases due to
the contamination of a close clump, for the same reason and at about
the same radius, the $<\rm{v}>$ profile should increase/decrease. As
for a decreasing profile, this might be likely due to the projection
effect of a few clumps centered around the center of the system and
having different mean velocities, i.e. somewhat aligned with the LOS
or, alternatively, of a large scale structure (LLS) elongated along
the LOS (e.g. a LLS filament).

The inspection of Fig.~\ref{figprofmul} shows that the $\sigma_{\rm
  v}$ profiles of peaks No. 1, 2 and 5 sharply increase with the
distance from the peak position.  Simultaneously, the $<\rm{v}>$
profiles decline (peak No. 1) or increase (peaks No. 2 and 5).  For
each of these clumps we attempt to detect the region likely not
contaminated by other clumps -- and thus reliable for kinematical
analysis -- as the region before the sharp increasing of the
$\sigma_{\rm v}$ profile (see the arrows in Fig.~\ref{figprofmul} for
peaks No. 1, 2, and 5). No conclusion can be driven for peaks No. 3
and 4 where the $\sigma_{\rm v}$ profile is decreasing.

\subsection{Substructure of the main system}
\label{main}

Here we present the results of our substructure analyses applied to
the 145 galaxies of the main system (MS), i.e. rejecting the galaxies
of the high velocity group (HVG) which might mask the real cluster
structure.

\begin{figure}
\centering 
\resizebox{\hsize}{!}{\includegraphics{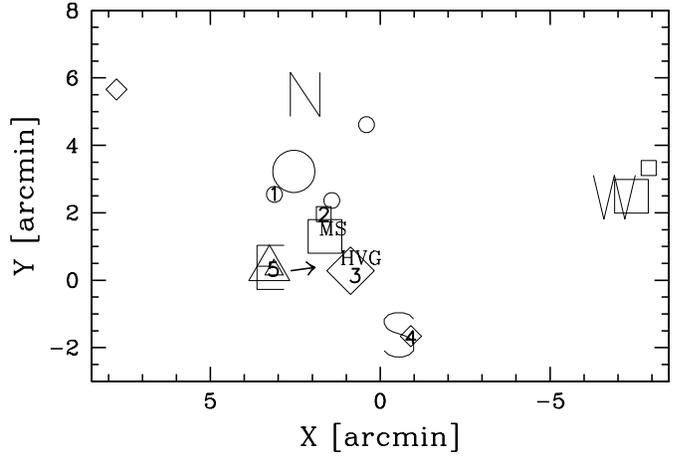}}
\vspace{-2cm}
\caption
{Summary of spatial distribution of cluster substructure.  Labels
  ``MS'' and ``HVG'' indicate the optical centers of the MS and of the
  HVG, respectively.  Other symbols refer to galaxy subclumps found in
  the MS using several, different approaches.  Triangles, squares,
  circles and rotated squares indicate 2D DEDICA peaks (large symbols)
  and luminous galaxies (small symbols) of the V1, V2, V3 and V4
  clumps. Labels ``N'', ``S'', ``E'' and ``W'' indicate the (biweight)
  centers of the four DS clumps.  Labels ``1'',
  ``2'', ``3'', ``4'' and ``5'' indicate the corresponding lensing
  mass peaks listed by M07: around peaks No. 1, 2, 4 and 5 we detect
  the P1, P2, P4 and P5 clumps.  On the base of the location of these
  subclumps and their velocities in Table~\ref{tabsub} we discuss the
  presence of five main structures (${\cal NE}1$, ${\cal NE}2$, ${\cal
    SW}$, ${\cal E}$ and ${\cal W}$ centered around the labels ``1'',
  ``2'', ``4'', ``5'' and ``W'', respectively) and of two minor
  structures (${\cal N}$ and ${\cal C}$ roughly located around the
  label ``N'' and the large rotated square, respectively), see
  Sect.~\ref{dyn}.  The arrow indicates the head tail radiogalaxy ID
  184 with the direction of its tail. The plot is centered on the
  X--ray cluster center.  }
\label{figsub}
\end{figure}

Table~\ref{tabsub} and Figure~\ref{figsub} summarize kinematical and
spatial properties of the subclumps we detect in the MS.  In
particular, Table~\ref{tabsub} lists for each clump the number of
galaxies (Col.~2); the mean velocity and its jacknife error (Col.~3);
the velocity dispersion and its bootstrap error (Col.~4); the luminous
galaxies contained within the analyzed clump (Col.~5); the name of the
corresponding structure discussed in Sect.~\ref{dyn} (Col.~6).  The
following subsections show the results recovered for each of the three
methods of analysis.

\begin{table}
        \caption[]{Results of the substructure analysis of the MS.}
         \label{tabsub}
                $$
         \begin{array}{l c l l l l}
            \hline
            \noalign{\smallskip}
            \hline
            \noalign{\smallskip}
\mathrm{Gal.Clump} & \mathrm{N_g} & \phantom{249}\mathrm{<v>}\phantom{249} & 
\phantom{24}\sigma_{\rm v}^{\mathrm{a}}\phantom{24}& 
\rm{Lum.gals}&\mathrm{Structures}\\
& &\phantom{249}\mathrm{km\ s^{-1}}\phantom{249} 
&\phantom{2}\mathrm{km\ s^{-1}}\phantom{24}& ID&\\
            \hline
            \noalign{\smallskip}
 
\mathrm{V1}^{\mathrm{b}}& 30&58568\pm56&300_{-46}^{+54}&205&{\cal E}\\
\mathrm{V2}^{\mathrm{b}}& 30&59455\pm18&\phantom{4}95_{-6}^{+6}&44,170&{\cal NE}2+{\cal W}\\
\mathrm{V3}^{\mathrm{b}}& 33&60113\pm19&106_{-13}^{+13}&204,160,137&{\cal NE}1\\
\mathrm{V4}^{\mathrm{b}}& 52&60920\pm33&238_{-36}^{+7}&106,264&{\cal SW}+{\cal C}\\
\mathrm{DS-N    }& 4 &58427\pm272& 415_{-415}^{+478}&-&{\cal N}\\
\mathrm{DS-S    }& 4 &61165\pm519& 791_{-471}^{+323}&106&{\cal SW}\\
\mathrm{DS-E    }& 7 &58686\pm292&682_{-239}^{+356}&205&{\cal E}\\
\mathrm{DS-W    }& 4 &59635\pm249& 380_{-112}^{+83}&-&{\cal W}\\
\mathrm{P\ 1}& 7 &60448\pm349& 811_{-71}^{+278}&204&{\cal NE}1\\
\mathrm{P\ 2}& 9 &59596\pm276& 749_{-88}^{+186}&170,160&{\cal NE}2\\
\mathrm{P\ 4}& 6 &61447\pm283& 579_{-151}^{+523}&106&{\cal SW}\\
\mathrm{P\ 5}& 6 &58634\pm325& 668_{-187}^{+570}&205&{\cal E}\\
              \noalign{\smallskip}
            \hline
            \noalign{\smallskip}
            \hline
         \end{array}
$$
\begin{list}{}{}  
\item[$^{\mathrm{a}}$] We use the biweight and the gapper estimators by 
Beers et
al. (1990) for samples with $\mathrm{N_g}\ge$ 15 and with
$\mathrm{N_g}<15$ galaxies, respectively (see also Girardi et
al. \cite{gir93}).
\item[$^{\mathrm{b}}$] 
The estimate of $\sigma_{\rm V}$ should
be considered a lower limit in these samples (see text).
\end{list}
 \end{table}

\subsubsection{Velocity distribution}
\label{veloms}

The velocity distribution of the MS is negatively skewed (at the c.l. of
$90-95\%$, skewness=$-0.330$) and light--tailed (at the c.l. of
$90-95\%$, kurtosis=$2.347$).  The W--test (Shapiro \& Wilk
\cite{sha65}) rejects the null hypothesis of a Gaussian parent
distribution at the $>99.9\%$ c.l..

\begin{figure}
\centering 
\resizebox{\hsize}{!}{\includegraphics{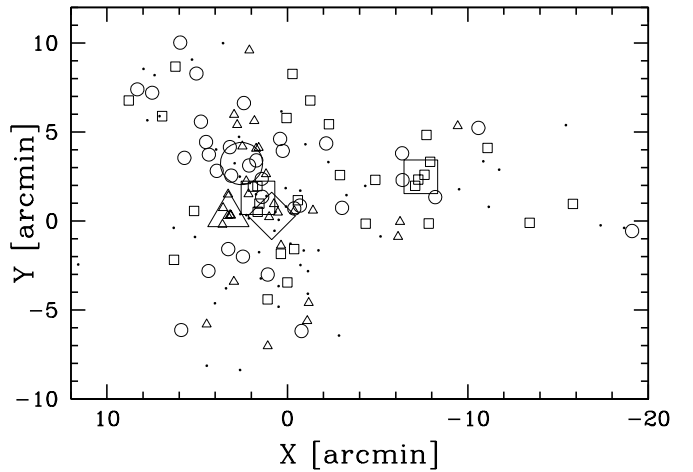}}
\vspace{-3cm}
\caption
{Spatial distribution on the sky of the 145 galaxies belonging to the
  MS showing the four groups recovered by the weighted gap
  analysis. Small triangles, squares, circles and dots indicate
  galaxies of the V1, V2, V3 and V4 groups, respectively. Large
  triangle, (2) squares, circle and rotated square indicate DEDICA
  peaks of the V1, V2, V3 and V4 galaxy distributions.}
\label{figkmmms}
\end{figure}

We detect three marginally significant gaps which divide the MS in
four groups of 30, 30, 33 and 52 galaxies (see Fig.~\ref{figvd} and
Table~\ref{tabgap}), hereafter defined as V1, V2, V3 and V4 from low
to high velocities. When compared two by two through the 2DKS--test,
these groups differ in spatial distribution: the V1 group differs both
from the V2 and V3 groups (at the $98\%$ and $94\%$ c.l.,
respectively); the V4 group differs from the V3 group (at the $94\%$
c.l.).  For each of the these groups Fig.~\ref{figkmmms} shows the
spatial distribution of galaxies and the corresponding peak according
to the 2D DEDICA procedure. For the V2 group, we find and plot two
peaks of comparable significance.  Figure~\ref{figsub} shows the
position of these peaks in relation to the lensing mass peaks listed
by M07.  Each of these groups contains one o more luminous galaxies
and, with the exception of V4, each group hosts one correspondent
luminous galaxy close to the respective peak in the galaxy
distribution (see Fig.~\ref{figsub}).  Properties of groups recovered
by kinematical analysis are listed in Table~\ref{tabsub}.  For the V1,
V2, V3 and V4 groups, the membership assignment might lead to an
artificial truncation of the tails of the distributions; thus the
values of velocity dispersion should be considered lower limits (e.g.,
Bird \cite{bir94}).

Using the results of the gap analysis to determine the first guess of
the KMM algorithm we find that a four-groups partition is a
significant better descriptor of the velocity distribution with
respect to a single Gaussian at the $99\%$ cl..  In particular, the
cluster partition is similar to that indicated by the above
weighted--gap analysis separating the MS in groups of 33,
27, 35 and 50 galaxies.

\subsubsection{Dressler--Shectman statistics}
\label{combms}

\begin{figure}
\centering 
\resizebox{\hsize}{!}{\includegraphics{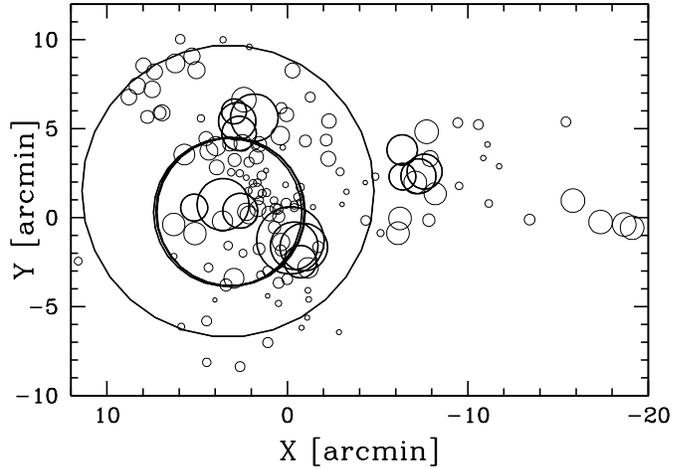}}
\vspace{-3cm}
\caption
{The result of Dressler-Shectman analysis 
as in Fig.~\ref{figds}, but for the 145 members of the MS.
  Here heavy circles indicate galaxies with $\delta_i > 1.95$ used to
  define the DS-N, DS-S, DS-E and  DS-W clumps (see text).}
\label{figdsms}
\end{figure}

The DS test on the MS gives a very marginal indication of substructure
(at the $91\%$ c.l.). However, as for the location of substructures,
the DS plot of the MS is similar to that recovered for the whole
sample (cf. Fig.~\ref{figds} and Fig.~\ref{figdsms}).  The main
difference is the disappearance of the northern substructure very
close to the X--ray cluster center which was likely due to the
galaxies of the HVG. The other three subclumps are still present (at
eastern, southern and distant western cluster regions). Moreover, the
analysis of the MS shows a northern subclump 5\arcm from the X--ray
cluster center. In order to better investigate the properties of DS
substructure we select the galaxies with the highest $\delta_i$ in
such a way that at least four galaxies can be assigned in an unique
way to each of the four DS subclumps. This leads to 19 galaxies with
$\delta_i>1.95$.  The kinematical properties of the four DS subclumps
are listed in Table~\ref{tabsub}. Figure~\ref{figsub} shows the
position of their (biweight) centers.

\subsubsection{Analysis of velocity dispersion profiles}
\label{2dms}

We also reanalyze the kinematical properties of galaxies surrounding
the lensing mass peaks listed by M07 (see Fig.~\ref{figprofmul} -
dashed lines).  We notice that for all the peaks where we have defined
a region likely uncontaminated by other clumps in Sect.~\ref{2d}
(peaks No. 1, 2 and 5), we confirm that these regions are devoid of
HVG galaxies.  Moreover, now we also find a sharp increase of the
$\sigma_{\rm v}$ profile for peak No. 4 and we define a likely non
contaminated region for the subsystem corresponding to this peak,
too. The groups formed by galaxies within the uncontaminated fiducial
regions are referred as P1, P2, P4 and P5 and their kinematical
properties are shown in Table~\ref{tabsub}.

In the case of peak No. 3 we still find a decreasing $\sigma_{\rm v}$
profile; thus we find no evidence for an individual, dynamically
important structure around this peak.

\section{Cluster structure and dynamics}
\label{dyn}

The value we find for the global velocity dispersion of the cluster
members, $\sigma_{\rm v}=1066_{-61}^{+67}$, is in agreement with
previous analyses (Carlberg et al. \cite{car96}; Borgani et
al. \cite{bor99}; Proust et al. \cite{pro00}; Mezzetti \& Girardi
\cite{gir01}). This value of the velocity dispersion is also
comparable to the average X--ray temperature assuming the
density--energy equipartition between gas and galaxies, i.e.
$\beta_{\rm spec}=1$
\footnote{$\beta_{\rm spec}=\sigma_{\rm v}^2/(kT/\mu m_{\rm p})$ with
  $\mu=0.58$ the mean molecular weight and $m_{\rm p}$ the proton
  mass.}, see Fig.~\ref{figprof} -- lower panel.

We analyze the cluster structure using the velocity distribution
analysis (weighted gap technique and KMM method), the
Dressler--Shectman statistics and the analysis of the velocity
dispersion profiles.  The structure of A520 is definitely very complex
and thus likely far from the dynamical equilibrium. The agreement
between $\sigma_{\rm v}$ and $T_{\rm X}$ might be due to an
enhancement of both quantities and, as already pointed out by M07,
gross properties and scaling relations are not always useful
indicators of the dynamical state of clusters.

Hereafter we analyze and discuss the structure of A520 starting from
the simplest hypothesis and then adding some degrees of complexity.

\subsection{Mass estimate}
\label{mass}

Making the usual assumptions (cluster sphericity, dynamical
equilibrium, that the galaxy distribution traces the mass
distribution), one can compute virial global quantities. Following the
prescriptions of Girardi \& Mezzetti (\cite{gir01}), we assume for the
radius of the quasi--virialized region $R_{\rm vir}=0.17\times
\sigma_{\rm v}/H(z) = 2.34$ \h -- see their eq.~1 after introducing
the scaling with $H(z)$ (see also eq.~ 8 of Carlberg et
al. \cite{car97} for $R_{200}$). Thus the cluster is sampled out --
although in a non--homogeneous way -- to $R_{200}$.  We compute the
virial mass (Limber \& Mathews \cite{lim60}; see also, e.g., Girardi
et al. \cite{gir98}):

\begin{equation}
M=3\pi/2 \cdot \sigma_{\rm v}^2 R_{\rm PV}/G-{\rm SPT},
\end{equation}

\noindent where SPT is the surface pressure term correction (The \&
White \cite{the86}), and $R_{\rm PV}$ is a projected radius (equal to
two times the projected harmonic radius).
%

The estimate of $\sigma_{\rm v}$ is robust when computed within a
large cluster region (see Fig.~\ref{figprof}).  The value of $R_{\rm
  PV}$ depends on the size of the sampled region and possibly on the
quality of the spatial sampling (e.g., whether the cluster is
uniformly sampled or not).  Considering the 155 galaxies within
$R_{\rm vir}$ we obtain $R_{\rm PV}=(1.72\pm0.09)$ \hh, where the
error is obtained via the jacknife procedure. The value of SPT
strongly depends on the radial component of the velocity dispersion at
the radius of the sampled region and could be obtained by analyzing
the (differential) velocity dispersion profile, although this procedure would
require several hundred galaxies. We decide to assume a $20\%$ SPT
correction as obtained in the literature by combining data on many
clusters sampled out to about $R_{\rm vir}$ (Carlberg et
al. \cite{car97}; Girardi et al. \cite{gir98}). We compute $M(<R_{\rm
  vir}=2.34 \hhh)=(1.7\pm0.2)$ \mquii.

Since the cluster center of A520 is not well defined and the spatial
sampling is not complete and homogeneous within $R_{\rm vir}$, one
could use an alternative estimate of $R_{\rm PV}$ on the basis of the
knowledge of the galaxy distribution. Following Girardi et
al. (\cite{gir98}; see also Girardi \& Mezzetti \cite{gir01}) we
assume a King--like distribution with parameters typical of
nearby/medium--redshift clusters: a core radius $R_{\rm c}=1/20\times
R_{\rm vir}$ and a slope--parameter $\beta_{\rm fit}=0.8$, i.e. the
volume galaxy density at large radii goes as $r^{-3
  \beta_{fit}}=r^{-2.4}$. We obtain $R_{\rm PV}=1.74$ \hh, where a
$25\%$ error is expected (Girardi et al. \cite{gir98}).  The mass
recovered by this method is then $M(<R_{\rm vir}= 2.34
\hhh)=(1.7\pm0.5)$ \mqui in excellent agreement with the above direct
estimate.

Our analysis of the cluster velocity distribution detects the presence
of a high velocity group (HGV) with a relative rest--frame  LOS velocity of
${\rm v_{\rm rf}}\sim 2000$ \ks with respect to the main system (MS), see
\S~\ref{velo}.  Therefore we might think that the cluster is better
described by the combination of the MS and the HVG, considered as two
separated entities. Assuming the dynamical equilibrium for both the MS
and the HVG we can compute independent virial radii and masses
$M(<R_{\rm vir}= 1.79 \hhh)=(8\pm2)$ \mqua and $M(<R_{\rm vir}= 0.74
\hhh)=(0.6_{-0.3}^{+0.8})$ \mquaa, respectively.

To compare the mass estimates derived for the two above cluster models
(relaxed cluster and the MS+HVG system) we consider the mass values
within 1 \h.  To rescale our mass estimates we assume that the system
is described by a King--like mass distribution (see above) or,
alternatively, a NFW profile where the mass--dependent concentration
parameter $c$ is taken from Navarro et al. (\cite{nav97}) and rescaled
by the factor $1+z$ (Bullock et al. \cite{bul01}; Dolag et
al. \cite{dol04}), i.e.  $c=4.16$, 4.96 and 6.18 for the whole
cluster, the MS and the HVG, respectively.  The relaxed cluster model
leads to a mass of $M(<1 \hhh)=(7.2-9.6)$\mqua while the addition of
the MS and the HVG masses leads to a mass of $M(<1 \hhh)=(4.0-6.8)$
\mqua [where the mass range includes a 1$\sigma$ error on the original
$M(<R_{\rm vir})$ estimate].

Using the above rescaling we can also compare our results with the
estimates recovered from X--ray and gravitational lensing analyses.
Lewis et al. (\cite{lew99}) used the ROSAT X--ray surface brightness
and ASCA temperature to estimate a mass of $M(<1.764 \hhh)=(11.3\pm1.1)$
\mqua (the ROSAT--PSPC estimate is converted in our cosmology).
X--ray mass is intermediate between our estimates since we obtain
$M(<1.764 \hhh)=(12-15)$ \mqua and $(6.1-10)$ \mqua for the two cluster
models. As for gravitational lensing, Dahle et al. (\cite{dah02})
obtained the projected mass $M_{\rm proj}(<1.111
\hhh)=11.7^{+3.9}_{-2.3}$ \mqua (see their Fig.~50 with conversion in
our cosmology). Our projected mass estimates for the two cluster
models are $M_{\rm proj}(<1.111 \hhh)=(12-18)$ \mqua and $(5.4-11)$ \mquaa,
where to make the projection we have considered that the cluster mass
distribution is truncated at one or at two virial radii. 

\subsection{Main system and high velocity group: relative dynamics}
\label{bim}

Continuing with the assumption of a cluster formed by the MS and the
HVG, we investigate their relative dynamics. We use different analytic
approaches which are based on an energy integral formalism in the
framework of locally flat spacetime and Newtonian gravity (e.g., Beers
et al. \cite{bee82}). The values of the relevant observable quantities
for the two--clumps system are: the relative LOS velocity in the rest
frame, ${\rm V_{\rm rf}}=2033$ \ks (as recovered from the MS and
the HVG); the projected linear distance between the two clumps,
$D=0.21$ \h (as recovered from optical centers of the MS and the HVG);
the mass of the system obtained by adding the masses of the two
subclusters each within its virial radius, log$M_{\rm
  sys}=14.9154_{-0.1512}^{+0.1264}$ (see Table~\ref{tabv}).

\begin{figure}
\centering
\resizebox{\hsize}{!}{\includegraphics{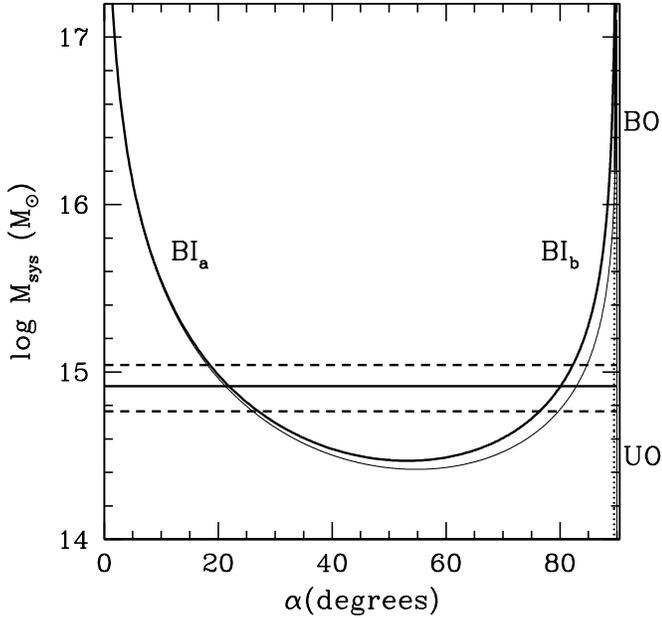}}
\caption
{System mass vs. projection angle for bound and unbound solutions 
  (solid and dotted curves, respectively) of the two--body model applied
  to the MS and the HVG subsystems.  Labels BI$_{\rm a}$ and BI$_{\rm
    b}$ indicate the bound and incoming, i.e. collapsing
  solutions (the main part of the solid curve).  Label BO indicates
  the bound outgoing, i.e. expanding solutions (the part of
  the solid curve which is roughly a vertical line).  Label UO
  indicates the unbound outgoing solutions (the dotted curve
  which is roughly a vertical line).  The horizontal lines give the
  observational values of the mass system and its uncertainties.  The
  bound and unbound regions according to the Newtonian criterion are
  indicated, too (above and below the thin curve, respectively).  }
\label{figbim}
\end{figure}

First, we consider the Newtonian criterion for gravitational binding
stated in terms of the observables as $V_{\rm r}^2D\leq2GM_{\rm
  sys}\rm{sin}^2\alpha\,\rm{cos}\alpha$, where $\alpha$ is the
projection angle between the plane of the sky and the line connecting
the centers of the two clumps. The thin curve in Fig.~\ref{figbim}
separates the bound and unbound regions according to the Newtonian
criterion (above and below the curve, respectively). Considering the
value of $M_{\rm sys}$, the MS+HVG system is bound between $21\degree$
and $83\degree$; the corresponding probability, computed considering
the solid angles (i.e., $\int^{83}_{21} {\rm cos}\,\alpha\,d\alpha$),
is 63\%.

Then, we apply the analytical two--body model introduced by Beers et
al. (\cite{bee82}) and Thompson (\cite{tho82}; see also Lubin et
al. \cite{lub98} for a recent application).  This model assumes radial
orbits for the clumps with no shear or net rotation of the
system. Furthermore, the clumps are assumed to start their evolution
at time $t_0=0$ with separation $d_0=0$, and are moving apart or
coming together for the first time in their history; i.e. we are
assuming that we are seeing the HVG prior to merging with the MS (at the time
t=11.022 Gyr at the cluster redshift, see Wright \cite{wri06}).  The
bimodal model solution gives the total system mass $M_{\rm sys}$ as a
function of $\alpha$ (e.g., Gregory \& Thompson \cite{gre84}).
Figure~\ref{figbim} compares the bimodal--model solutions with the
observed mass of the system.  The present bound outgoing solutions
(i.e. expanding), BO, are clearly inconsistent with the observed mass.
The possible solutions span these cases: the bound and present
incoming solution (i.e. collapsing), BI$_{\rm a}$ and BI$_{\rm b}$,
and the unbound--outgoing solution, UO.  For the incoming case there
are two solutions because of the ambiguity in the projection angle
$\alpha$.  We compute the probabilities associated to each solution
assuming that the region of $M_{\rm sys}$ values between the
uncertainties are equally probable for individual solutions: $P_{\rm
  BIa}\sim88$\%, $P_{\rm BIb}\sim12$\%, $P_{\rm UO}\sim 7\times
10^{-6}$\%.  Thus it is very likely that the HVG lies in front of the
cluster just infalling onto it.

Notice, however, that the centers of the HVG and the MS are not well
determined.  The HVG has too small a number of galaxies for a precise
center determination. As for the MS center, we might adopt the X--ray
cluster center instead of the optical cluster center.  These
uncertainties do not change the bulk of our results.  For instance, in
the case where we assume that the X--ray center is the MS center, i.e.
a smaller projected linear distance between the two clumps ($D=0.1$
\hh), the effect is to increase the boundary probability (at $75\%$
c.l. for the Newtonian model) and to yield more extreme values for the
bound solutions ($\alpha \sim 15$ and 85 degrees).  Also, possible
underestimates of the masses (e.g., if the MS and the HVG actually
extend outside of the virial radii we estimate for them) would lead to
binding probabilities larger than those computed above, as well as
more extreme values for $\alpha$.  Thus the analysis here displayed
should be considered a lower limit for our conclusions in
\S~\ref{peak3}, where we propose the existence of a cluster accretion
along the LOS (i.e. $\alpha$ close to 90 degrees).

\subsection{NE--SW merger}
\label{nesw}

Although the presence of the HVG system is maybe the most important
for the optical virial mass computation, we find that A520 shows a
much more complex structure. In fact, we detect several subsystems in
the MS along the NE--SW and E--W directions (see Figs.~\ref{figradio}
and \ref{figsub}).

Clumps around lensing mass peaks No. 1, 2 and 4 of M07 define the
direction of the likely merger along the NE--SW direction with the SW
structure having crossed the NE structure (Markevitch et
al. \cite{mar05}).

Around peaks No. 1 and 2 we detect two structures (hereafter ${\cal
  NE}1$ and ${\cal NE}2$) using the analysis of the velocity
distribution (V3 and V2) and velocity dispersion profiles (P1 and
P2). These structures have a relative rest--frame LOS velocity of
$\sim 700$ \kss. Each structure hosts, close to the center, a luminous
galaxy having typical velocity of the structure (IDs 204 and 170 in
${\cal NE}1$ and ${\cal NE}2$, respectively).

Structures ${\cal NE}1$ and ${\cal NE}2$ are so close in space and in
velocity that the ``uncontaminated'' regions detected by $\sigma_{\rm
  v}$ profiles are slightly superimposed (see
Fig.~\ref{figradio}). Moreover, at $\lesssim 0.5$ \arcmin from ID~170
(the ${\cal NE}2$ central, luminous galaxy) we find the luminous
galaxy ID~160 which belongs to V3 group, i.e. to the ${\cal
  NE}1$ structure.  Such close couples of galaxies having different
velocities are often observed in clusters (Boschin et
al. \cite{bos06}; Barrena et al. \cite{bar07a}) and are the likely
tracers of a previous cluster merger. Indeed cluster merger is thought
to be the cause of the formation of dumbbell galaxies (e.g., Beers et
al. \cite{bee92}; Flores et al. \cite{flo00}).  Therefore ${\cal
  NE}1$+${\cal NE}2$ is likely to form a single, although not yet well
relaxed, structure and represent the real, original main cluster. In
fact, the combined velocity of ${\cal NE}1$ and ${\cal NE}2$ is $\sim
60000$ \kss, similar to the mean velocity of the MS, and the ${\cal
  NE}2$ position is close to the optical center of the MS.  The
dynamical importance of the ${\cal NE}1$+${\cal NE}2$ structure
explains why the SW structure has been reported to have suffered
significant damage in the merging, as shown by the pieces of the
cluster core detected in X--ray (Markevitch et al. \cite{mar05}).

Around peak No. 4 we detect a structure (hereafter ${\cal SW}$) both
using DS analysis (DS-S) and studying the velocity dispersion profile
(P4). It is characterized by a high velocity ${\rm v}\sim 61300$
\kss, i.e. ${\rm v_{\rm rf}}\sim +1100$ \ks from the ${\cal
  NE}1$+${\cal NE}2$ complex.  ${\cal SW}$ is not individually
detected in the velocity distribution. However, since it hosts close
to its center one of the two luminous galaxies of the V4 group
(ID 106 with ${\rm v}=61277$ \kss), ${\cal SW}$ is likely a part of
V4 (see also discussion in Sect.~\ref{peak3}).

A possible structure related to the NE--SW merger is DS-N (hereafter
${\cal N}$) having a small velocity ${\rm v}\sim 58400$ \kss. It is
only detected through the Dressler-Shectman analysis and does not
contain a luminous galaxy; therefore we neglect it in the following
discussion.  Notice however that it roughly corresponds to the N peak
in the lensing mass map of Okabe \& Umetsu (\cite{oka08}).

\begin{figure}
\centering
\resizebox{\hsize}{!}{\includegraphics{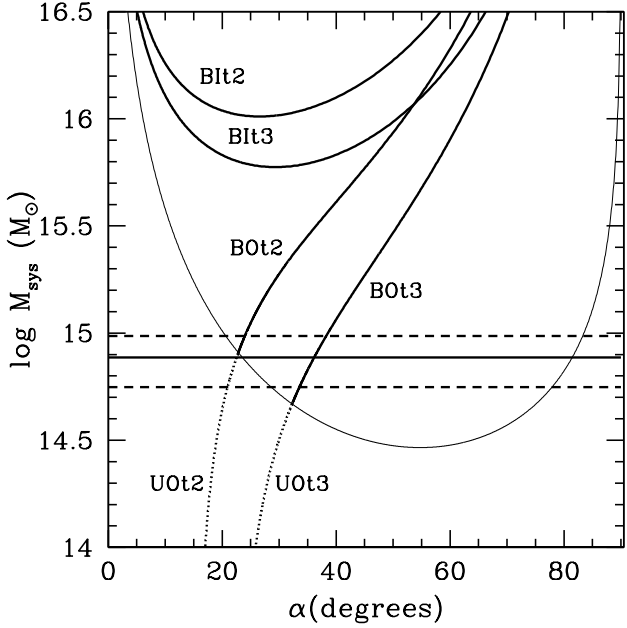}}
\caption
{System mass vs. projection angle for bound outgoing and unbound
  solutions (solid and dotted curves, respectively) of the two--body
  model applied to the ${\cal NE}1$+${\cal NE}2$ and ${\cal SW}$
  subsystems.  Labels BIt2, BOt2 and UOt2 indicate the curves
  corresponding to bound incoming, bound outgoing and unbound outgoing
  solutions assuming that the system is observed at the time $t=$0.2
  Gys after the core crossing. Labels BIt3, BOt3 and UOt3 have the
  same meaning but assuming that the system is observed at $t=$0.3 Gys
  after the core crossing.  The horizontal lines give the
  observational values of the mass system and its uncertainties.  The
  bound and unbound regions according to the Newtonian criterion are
  indicated, too (above and below the thin curve, respectively).  }
\label{figbimx}
\end{figure}

To attempt a more detailed analysis of the NE--SW merger we apply the
bimodal model considering the interaction between ${\cal NE}1$+${\cal
  NE}2$ (likely corresponding to the main part of the MS, see above) and
${\cal SW}$ assuming we are looking at them after their core crossing as
suggested from X--ray data. As parameters of the model we use a
relative LOS velocity ${\rm V_{\rm rf}}=1100$ \ks and a relative projected
distance $D=0.8$ \h.  We assume the MS mass as the mass for the whole
system. In order to apply the two--body model we assume that the time
$t_0=0$ with separation $d_0=0$ is the time of their core crossing and
that we are seeing the cluster a few $10^8$ years after the merging.
In fact, a few $10^8$ years is the time scale in which the
relativistic electrons lose energy, i.e. the lifetime of the radio
halo (e.g., Giovannini \& Feretti \cite{gio02}).  Figure~\ref{figbimx}
shows the results for a time of $t=$0.2 and 0.3 Gyrs after the core
crossing.  The likely bound solution is then an outgoing one, i.e. the
SW structure is now behind the NE structure going away from it.  In
particular, an angle of $\alpha\sim 30$ leads to an outgoing velocity
of $\sim 2200$ \kss.

The Mach number of the shock is ${\cal M}={\rm v}_{\rm s}/c_{\rm s}$,
where ${\rm v}_{\rm s}$ is the velocity of the shock and $c_{\rm s}$ in
the sound speed in the pre--shock gas (see e.g., Sarazin \cite{sar02}
for a review). In the stationary regime we can assume that ${\rm
  v}_{\rm s}$ is the merger velocity $2200$ \kss.  Assuming the
equipartition of energy density between gas and galaxies and using the
$\sigma_{\rm v}$ of the MS we obtain $c_{\rm s}\sim 812$ \ks from the
thermal velocity.  Therefore we estimate ${\cal M}\sim 2.7$, which is
in reasonable agreement with ${\cal M}= 2.1^{+0.4}_{-0.3}$ and ${\cal
  M}= 2.2^{+0.9}_{-0.5}$ recovered from X--ray data (Markevitch et
al. \cite{mar05}).

Notice that $t=0.2-0.3$ Gyrs and $\alpha \sim 30$ degrees are larger,
but comparable, to the merger parameters recovered by Markevitch et
al. (\cite{mar02}) for a similar scenario in the cluster 1E0657--56
which hosts a radio halo and shows a bow shock, too (0.1--0.2 Gyrs and
10--15 degrees). Indeed, this agreement is not casual since as noticed
by Markevitch et al. (\cite{mar05}) to observe a shock front ``one has
to catch a merger at a very specific stage when the shock has not yet
moved to the outer, low surface brightness regions and at a
sufficiently small angle from the sky plane, so that projection does
not hide the density edge''.

\subsection{Accretion along the E--W direction}
\label{seenww}

M07 suggested a possible secondary E--W merger related to lensing mass
peaks No. 3 and 5. However, these authors doubted  the dynamical
importance of the structure around the peak No. 5 since this mass peak is
poorly significant and the mass--to--light ratio is quite low (but see
Okabe \& Umetsu \cite{oka08} where the corresponding C3 peak is
quite significant).

We find a strong dynamical evidence of a structure around peak No. 5
(hereafter ${\cal E}$) at ${\rm v}\sim 58600$ \ks (i.e.  ${\rm v_{\rm
    rf}}\sim -1150$ \ks from the ${\cal NE}1$+${\cal NE}2$ complex) using
the velocity distribution analysis (V1), the velocity dispersion
profile (P5) and DS analysis (DS-E). Moreover, ${\cal E}$ hosts its
luminous galaxy ID 205 close to its center.  Comparing ${\cal E}$
with ${\cal SW}$ we notice that ${\cal E}$ lies at a similar velocity
distance and at much smaller spatial distance with respect to the
${\cal NE}1$+${\cal NE}2$ complex. However, since there is no sign of
a strong, present interaction from X--ray data, we suspect that ${\cal
  E}$ might be a high--speed remnant of a previous merger.

We also detect a structure (hereafter ${\cal W}$) located in the
western external region not sampled by previous gravitational lensing
analyses.  This structure is found using both the velocity
distribution analysis (part of V2) and the DS analysis (DS-W) and is
characterized by a velocity ${\rm v}\sim 59600$ \ks (i.e.  ${\rm
  v_{\rm rf}}\sim -300$ \ks from the ${\cal NE}1$+${\cal NE}2$
complex).  It hosts a luminous galaxy close to its center, too
(ID~44). Since the ${\cal W}$ velocity is similar to that of the
${\cal NE}1$+${\cal NE}2$ complex and the projected spatial distance
is about 2 \hh, ${\cal W}$ might be a distant subclump well far from
the merging, infall phase.

In conclusion we strongly reinforce the possibility of an accretion onto
A520 along the E--W direction.

\subsection{Nature of the ``massive dark core''}
\label{peak3}

Finally we discuss the region around lensing mass peak No. 3 for which M07
found a very large mass--to--light ratio claiming for the presence of
a massive dark core (but see Okabe \& Umetsu \cite{oka08} where the
corresponding C1 peak is not particularly pronounced).

This peak is the only M07 peak for which our analysis of the velocity
dispersion profile does not support the presence of an individual
structure.  The peak of the V4 galaxy distribution is close to peak
No. 3, but the V4 group is the only one which does not host any
luminous galaxy close to its center. In fact, the two V4 luminous
galaxies are located well far in the northern and southern cluster
regions (see Fig.~\ref{figsub}). Moreover, several galaxies of V4 are
likely to be associated with the V4 southern luminous galaxy,
i.e. with the ${\cal SW}$ structure we discuss in
Sect.~\ref{nesw}. Therefore, only part of galaxies we assign to V4 are
likely really connected with the region around peak No. 3
(hereafter we name this minor structure ${\cal C}$).

In conclusion, the existence of an individual, very important
structure associated to peak No. 3 is not supported by our kinematical
analysis. Rather, we find evidence of two groups centered in that
region: the minor  ${\cal C}$ clump  (at v$\sim 60900$ \kss)
and the HVG (at v$\sim 62400$ \kss) suggesting the accretion onto the
cluster along the LOS. In agreement with this idea, in the MS+HVG
bimodal model of Sect.~\ref{bim} we prefer the bound solution with
$\alpha \gtrsim 80 $ degrees, with the HVG almost LOS aligned with the MS and
infalling onto it. A scenario of a few groups at different velocities
agrees with the high velocity dispersion of galaxies we measure around
this peak, see also the high velocity ridge in the merging cluster
Abell 521 (Ferrari et al. \cite{fer03}).

\begin{figure}
\centering 
\resizebox{\hsize}{!}{\includegraphics{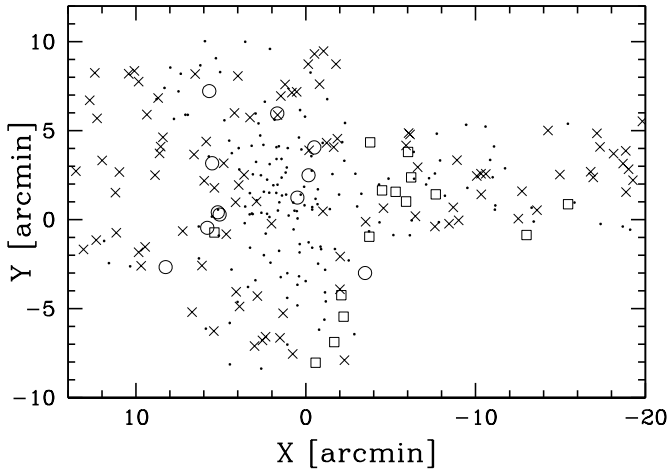}}
\vspace{-3cm}
\caption
{Spatial distribution on the sky of the 293 galaxies having redshifts
  in the cluster field.  Dots
  and large symbols indicate member and non member galaxies,
  respectively. In particular, squares and circles indicate galaxies
  belonging to back1 and back2 groups.}
\label{figback}
\end{figure}

Working in this scenario we also look for other possible groups
aligned with the LOS.  The cluster peak is not well isolated in the
velocity space in its high velocity limit, where a few small peaks are
present. We analyze the two closest groups of non--member galaxies,
hereafter back1 and back2, formed by 15 and 11 galaxies, respectively
(see Fig.~\ref{figvd}).  Figure~\ref{figback} shows the spatial
distribution of these ``background'' groups.  Galaxies of back1 and
back2 have a spatial distribution different from other non--member
galaxies (at the $99.96\%$ c.l. $95\%$ c.l.  according to
2DKS--test). While galaxies of back1 lie at the SW edges of the
sampled field and seem to have no connection with the cluster,
galaxies of back2 are loosely distributed in central-intermediate
cluster regions. Moreover, back2 group is characterized by a
remarkably small velocity dispersion $\sigma_{\rm v}\sim 150$ \ks and
a distance of $\Delta z \sim 0.02$ from A520.  Thus back2 might be a
very loose group connected with the cluster, since LSS connections are
likely found between systems separated by $\Delta z\sim 0.02$ (e.g.,
Arnaud et al. \cite{arn00}).

Finally, we discuss the head tail radio galaxy (ID 184). According to
Bliton et al (\cite{bli98}) this class of radio galaxies could be
related with bulk motion of the intergalactic medium and might
indicate the presence of a merger. ID 184 has the tail oriented
opposite to the cluster center as expected in a radiogalaxy infalling
onto the cluster (see Figs. \ref{figradio} and \ref{figsub}).  Having
high velocity v$\sim 60972$ \kss, ID 184 is likely not a galaxy of the
low velocity structure ${\cal E}$, rather it might be connected to the
infall along the LOS.

\section{Conclusions: A520 at the crossing of three LSS filaments}
\label{concl}

Our findings agree with a scenario where A520 is forming at the
crossing of three filaments of the LSS: the NE--SW one, the E--W one,
the one along about the LOS. Clusters are expected to form through the
accretion along three main filaments according to the LSS formation in
the CDM scenario (Shandarin and Zeldovich \cite{sha89}, see also
beautiful images of simulated clusters Springel et
al. \cite{spr05}\footnote{see
  http://www.mpa-garching.mpg.de/galform/millennium/}). Indeed, a few
examples of clusters forming at the crossing of two filaments were
already observed (e.g., Arnaud et al. \cite{arn00}; Cortese et
al. \cite{cor04}; Boschin et al. \cite {bos04}; Braglia et
al. \cite{bra07}) -- see also Matsuda et al. (\cite{mat05}) for a
protocluster at the crossing of three filaments.

In this scenario the massive dark core found  by M07 analysis would
coincide with the peak of the collisional component as shown by X--ray
data only due to the particular angle of view of the observer. In
fact, the X--ray peak likely traces the potential well of the forming
cluster, while the filament aligned with the LOS, projected onto the
cluster center, would produce the peak in the 2D mass distribution.
Indeed, the hypothesis of a LSS filament projected onto the location
of peak No. 3 was already suggested by M07 since is not in obvious
contrast with their gravitational lensing data and X--ray data.

Our analysis shows how powerful is the study of the internal cluster
dynamics on the base of velocities and positions of member galaxies.
It provides additional information which  complements X--ray and
gravitational lensing analyses. Other insights into A520 might be
recovered from the knowledge of galaxy properties (see e.g. Ferrari et
al. \cite{fer03}; Boschin et al. \cite{bos04}). In particular,
important information comes from the spectral types of member
galaxies, since star formation could increase or, alternatively, stop
during the merging phase; thus the spectral signatures of past
activity are useful to determine the relevant time--scales (e.g.,
Bekki \cite{bek99}; Terlevich et al. \cite{ter99}). We are planning
further studies of A520 in this perspective.

\begin{acknowledgements}
We would like to thank Federica Govoni for the VLA radio image and
Maxim Markevitch for the Chandra X-ray image they kindly provided
us. We also thank Stefano Borgani for useful discussions.

This publication is based on observations made on the island of La
Palma with the Italian Telescopio Nazionale Galileo (TNG), operated by
the Fundaci\'on Galileo Galilei -- INAF (Istituto Nazionale di
Astrofisica), and with the Isaac Newton Telescope (INT), operated by
the Isaac Newton Group (ING), in the Spanish Observatorio of the Roque
de Los Muchachos of the Instituto de Astrofisica de Canarias.

This work was partially supported by a grant from the Istituto
Nazionale di Astrofisica (INAF, grant PRIN--INAF2006 CRA ref number
1.06.09.06).

\end{acknowledgements}

\end{document}